\documentclass[11pt,review,1p,number,sort&compress,a4paper]{elsarticle}
\usepackage{graphicx}
\usepackage{lscape}
\usepackage{longtable,array}
\usepackage{amsmath}

\usepackage[dvipsnames]{xcolor}
\graphicspath{{/}}
\DeclareGraphicsExtensions{.pdf,.png,.jpg,}

\newcommand{\etal}{{\it et al.} }
\newcommand{\ai}{{\it ab initio}}
\newcommand{\cm}{cm$^{-1}$}

\newcommand{\aipes}{{\it ab~initio}~PES}

\begin{document}

\title{A new spectroscopically-determined potential energy surface and \emph{ab initio} 
dipole moment surface for high accuracy HCN intensity calculations}

\author[iapras]{Vladimir  Yu. Makhnev}

\author[iapras]{Aleksandra A. Kyuberis}

\author[iapras,ucl]{Oleg L. Polyansky}

\author[iapras]{Irina I. Mizus}

\author[ucl]{Jonathan Tennyson\corref{cor1}}
\ead{j.tennyson@ucl.ac.uk}

\author[iapras]{Nikolai F. Zobov}

\cortext[cor1]{Corresponding author}

\address[iapras]{Institute of Applied Physics, Russian Academy of Science,
Ulyanov Street 46, Nizhny Novgorod, Russia 603950}
\address[ucl]{Department of Physics and Astronomy, University College London,
Gower Street, London WC1E 6BT, United Kingdom}
\date{\today}

\begin{abstract}

\label{sec.abst}
Calculations of transition intensities for small molecules like
H$_2$O, CO, CO$_2$ based on s high-quality potential energy surface
(PES) and dipole moment surface (DMS) can nowadays reach sub-percent
accuracy.  An extension of this treatment to a system with more
complicated internal structure -- HCN/HNC (hydrogen cyanide/hydrogen
isocyanide) is presented.  A highly accurate
spectroscopically-determined PES is built based on a recent \aipes\ of
the HCN/HNC isomerizing system.  588 levels of HCN with
$J$~=~(0,~2,~5,~9,~10) are reproduced with a standard deviation from
the experimental values of $\sigma=0.0373$ \cm\ and 101 HNC levels
with $J$~=~(0,~2) are reproduced with $\sigma=0.37$ \cm. The
dependence of the HCN rovibrational transition intensities on the PES
is tested for the wavenumbers below 7200 \cm.  Intensities are
computed using wavefunctions generated from an \ai\ and our optimized
PES.  These intensities differ from each other by more
than 1\%\ for about 11\% of the transitions tested, showing
the need to use an optimized PES to obtain wavefunctions for
high-accuracy predictions of transition intensities.  An \ai\
  DMS is computed for HCN geometries
lying below 11~200 \cm. Intensities for
HCN transitions are calculated using a new fitted PES and newly
calculated DMS.  The resulting intensities compare much better with
experiment than previous calculations.  In particular, intensities of
the H--C stretching and bending fundamental transitions are predicted
with the subpercent accuracy.

\end{abstract}

\maketitle

\section{Introduction}
\label{sec.intro}

The isomerizing system HCN/HNC is important in several areas of science. 
%The reasons for this lies in the combination of the properties of the system, namely:
%\begin{enumerate}
%\item HCN is a stable molecule with a metastable isomer - HNC, as a result of which this system is an extremely flexible and highly anharmonic;
%\item But from the other point of view, HCN is a linear small molecule, available for a comprehensive theoretical study;
%\item Properties of equilibrium states of isomers and the isomerization barrier of the system distinguish it from among other systems with isomerization. 
%\end{enumerate}
One significant application of studies HCN/HNC spectra is for
astrophysics.  Both isomers have been detected in space objects of
various types \cite{78RiCaHa.HCN,74UlCoxx.CH3CN,84ErGuJo.HCN}, often with
abundances far from thermodynamic equilibrium. Observation of both HCN
and HNC is facilitated by the fact that both molecules have non-zero
dipole moments and thus can be detected by observing pure rotation
spectral lines in the submillimeter range or
vibrational-rotational lines in the infrared. Along with CO and
H$_2$O, HCN is one of the few molecules that have been detected in the infrared
in protoplanetary disks \cite{14BruHarDis.HCN.proto}. 
In the last two years,  HCN/HNC 
 has been identified in various
astronomical environments including exoplanets \cite{jt629},
satellites \cite{16MolNixCor.HCN.titan}, comets
\cite{16WirLerKal.HCN.comet}, and planets \cite{17LeGuBu.HCN}. Modern
observing techniques
impose stricter requirements on the accuracy of the available
theoretical data \cite{14deBiBr}, namely, transition frequencies and intensities.
%Recently, as part of the analysis of the atmosphere of the exoplanet \cite{jt629} from the constellation of Cancer E, the HCN compound was first discovered in the atmosphere of this planet. Before that, HCN was the first molecule used to demonstrate the necessary of comprehensive linelists on atmospheric models of cool stars \cite{84ErGuJo.HCN}. Also HCN-HNC system is an important constituent of protoplanetary disks [ \textbf{Ro-vibrational excitation of an organic molecule (HCN) in protoplanetary disks, Bruderer, Simon; Harsono, Daniel; van Dishoeck, Ewine F., A\&A, 2015}], satellites [\textbf{E. S. Wirström, M. S. Lerner, P. Källström, A. Levinsson, A. Olivefors and E. Tegehall, A\&A, 588 (2016)}]  interstellar matter and comets and plays an important role in providing opacity in carbon-rich stars \cite{jt321}. 
%\cite{hcnVoy1981,Titan1978,Comet1974,jt631}.

In addition to astrophysical studies, the HCN/HNC system is of
importance in the quantum chemical analysis. As noted by Mellau \cite{10Mexxxa.HNC}, 
the potential energy surface of [H,C,N], in
contrast to that of the [H,C,P] system, results in wavefunctions of the two
isomers H--CN and CN--H, located in two different minima merging step
by step to a single delocalized wavefunction where only a single
``combined'' H$_{0.5}$-CN-H$_{0.5}$ molecule exists. 
Baraban \etal~\cite{15BaChMe.HCN} and Mellau \etal~\cite{16MeKyPo.HCN}
studied the behavior of this system in the saddle point region.
In particular, Mellau \etal~\cite{16MeKyPo.HCN}
showed that if the saddle point-localized states of the [H,C,N] system
can be detected in molecular spectra, they would allow one to define
transition states without the need for any quantum frequency
analysis.

The various applications of the HCN/HNC spectra require further spectroscopic
studies of this molecule. The ability to predict  line positions
in different frequency ranges and higher temperatures is required for
the analysis of the experimental observations of the HCN spectra.
In particular it has been suggested that the ratio of HCN to HNC could
act as a thermometer in hot atmospheres such as those of cool carbon stars
\cite{jt304}.

Last but not least, the rovibrational spectrum of HCN has unique
features that, thus far, have provided difficult to describe
theoretically.  Maki \etal~\cite{95MaQuKl.HCN} measured
the C$\equiv$N stretching fundamental band and analyzed it in terms of
transition dipole moments and Herman-Wallis constants. The unusual
intensity patterns for the P- and R-branch of this $(\nu_{1}\nu_{2}^{l}\nu_{3})=(0 0^0 1)$ band
were explained by large rotation-vibrational interactions meaning that
the band dipole was significantly less than permanent dipole
moment and, in particular, the transition dipole moment changed sign
at a non-zero value of the rotational quantum number,  $J$, 
for almost all isotopologues. The observed
band dipole and Herman-Wallis factor values have not so far been
accurately reproduced theoretically. In particular, calculations by
van Mourik \etal~\cite{jt273} gave band dipoles which are
more than 30 \%\ larger the observed ones. Conversely the
dipole values computed by Botschwina \etal~\cite{97Bot.theory} agree
well with experiment but their Herman-Wallis factors do not.

%spectroscopic needs. There was also mentioned by Mellau \etal \cite{16MeKyPo.HCN} that 
%for the future experimental studies there would be needed a new, consistent and very accurate 
%data sets: states files and linelist based on \ai\ calculations. 

Lodi and Tennyson \cite{jt522} presented a computational
strategy to produce accurate line intensities and associated uncertainties.
Among other things, the strategy includes computing two sets of
nuclear-motion wavefunctions and energy levels using two different,
high-quality, PESs. This approach has been employed successfully for
a number of studies \cite{jt687,jt625,jt690}. A reliable \ai\ PES
is already available for the HCN/HNC system \cite{jt689} and
the present work  provides the second PES needed for such calculations. 

%\red{From ``Accurate line intensities of methane from first-principles calculations'' by Nikitin Rey Tyuterev:}

%\textit{``A common issue for global variational spectra predictions is a very high sensitivity of line intensities to sharp vibration-rotation resonances. This problem occurs for all polyatomic molecules but is particularly pronounced for high-energy states and dense energy level patterns. Because of accidental coincidence of vibration-rotation levels the Coriolis and anharmonic couplings could produce a strong mixing of the basis set functions resulting to well known ``intensity borrowing'' effects among weak and strong bands. Such transitions are usually referred to as ``unstable'' lines or ``resonance'' lines. Their intensities are not determined by the proper dipole transition moments of the corresponding bands but critically depend on the energy distances between the resonance partners. In such cases a tiny variation of the PES parameters could result in very big changes of instable line intensity even if the same DMS was used in the calculations. In other words, the transition probabilities $R_{nm}(2)$ of these lines are anomalously sensitive to the resonance mixing of wavefunctions and not to the dipole moment. Consequently, such unstable line intensities should not be included in the validation of the ab initio DMS.''}

%\red{I see this as a key part of explanation of need of 2nd PES (after \ai\ one)}

As the \aipes\ for the HCN/HNC system computed by van Mourik~\etal
\cite{jt273} does not satisfy modern spectroscopic needs, we recently
constructed a new one \cite{jt689}. Despite significantly increasing
the accuracy of the predicted rovibrational energy levels,
calculations with this PES still do not reach the 0.1 \cm\ level of
accuracy which places most computed line positions within their observed
line width in practical spectra.  Our experience with the water \cite{jt438,jt550} and
ammonia \cite{jt634} molecules shows that even a small correction to a
high-quality \aipes\ can give us a notable improvement in the
resulting energy levels.  In this context, we note that the previous
empirical PES constructed by Varandas and Rodrigues
\cite{06VaRoxx.HCN} is actually less accurate than our
\ai\ one \cite{jt689}.

This work is dedicated to the semi-empirical optimization of the PES
starting from our recent \aipes. The goal of such optimization is twofold. First, the
most accurate line positions will be obtained if the optimized PES is
used for the rovibrational energy levels calculations.  Second, the
use of an optimized PES will provide accurate wavefunctions, which
can be subsequently used in the high accuracy line intensity
calculations.  However, to achieve the high accuracy
intensity calculations, accurate wavefunctions are not enough. A
new, more accurate dipole moment surface (DMS) is also necessary.
Such a DMS is also presented here.

The paper is organized as follows. Section \ref{sec.base} gives the
technical details of our \ai\ calculations and some details of the
\ai\ energies fit. Section \ref{sec.fit} presents the fit of
vibrational and rotation-vibrational levels of HCN and HNC; comparisons
with the known experimental and \ai\ energy levels are also presented.
Sector \ref{sec.dip_cons} contains the details of dipole moment
surface construction. Section \ref{sec.intens} provides an intensity
analysis for the resulting PES and DMS. In section \ref{sec.diss} we
discuss our results and directions of the future work.

\section{Review of  existing surfaces}
\label{sec.base}
\subsection{Basic \ai\ potential energy surface}

In a recent paper \cite{jt689}, henceforth referred to as I, we
constructed a new \ai\ potential energy surface (\aipes).  Our
strategy was similar in spirit with various ``model chemistry''
schemes used in theoretical thermochemistry \cite{HEAT-paper} and with
the focal-point analysis \cite{focal-point}; it was based on a main,
high quality PES, to which several correction surfaces were added.  We
previously used this approach for the water \cite{jt550}, H$_2$F$^+$
\cite{h2fp} and ammonia \cite{jt634} molecules. But, unlike these
ten-electron systems, the electronic structure of [H,C,N] is more
complicated because of a multiple (triple) bond, which involves six
bonding electrons instead of the usual two ones in a covalent single
bond. For water molecule, we achieved 0.1 \cm\ accuracy \ai\
for the rovibrational energies \cite{jt550}. In the case of HCN/HNC
system, calculations with the \aipes\ of I \cite{jt689} reproduced the
extensive set of experimental data below 7000 \cm\ with a
root-mean-square deviation $\sigma = 0.37$ \cm\ when a semi-empirical
non-adiabatic correction was applied and $\sigma = 1.5$ \cm\ for
levels below 25000 \cm\ using a purely \ai\ approach.
%Such values of standard deviation suggest that our success with water \cite{jt550} 
%could be successfully (in smaller energy range) achieved for molecules with two second-row atoms and
%containing more electrons than water. 
%\blue{Why? This is orders of magnitude
%worse than the water result. The statement needs justifying.}\green{I see the sense as the whole paragraph.}

In I we used the functional form suggested by van Mourik~\etal \cite{jt273} as an analytical representation of the whole
\aipes\ in Jacobi coordinates:
\begin{equation} \label{main_func_form}
V_{ai}(r,R,\gamma) = \sum_{i,j,k}{A_{ijk}X^i(R,r,\gamma)Y^j(r,\gamma)P^k(\cos{\gamma})},
\end{equation}
where $r$ is the CN, $R$ is the ``H -- center-of-mass of CN'' distance  and $\gamma$ is the angle between $r$ and $R$, measured in radians, with $\gamma = 0$ representing HNC and HCN at $\gamma = \pi$. 
$X$ and $Y$ are functions of primarily the $R$ and $r$ coordinates, respectively, and are based on a Morse coordinate transformation, with
\begin{equation}
X(R, r, \gamma) = 1 - e^{-\alpha_R(\gamma)[R - R_e(\gamma, r)]},
\end{equation}
\begin{equation}
Y(r, \gamma)=1 - e^{-\alpha_r[r - r_e(\gamma)]}. 
\end{equation}
A total of 277 \aipes\ parameters were determined from the fit to 1541
aug-cc-pCV6Z multi-reference configuration interaction (MRCI) \ai\
points.  More details about the functional form are given by van
Mourik \etal~\cite{jt273} and about the \ai\ calculations in I.

\subsection{The dipole moment surface}

In the work by van Mourik \etal~\cite{jt273} dipole moments were
calculated at the selected 242 grid points (for both HCN and HNC
wells) employing cc-pCVQZ CCSD(T) wavefunctions with all electrons
correlated. At each grid point dipole moment components were computed
via the finite-field method. This DMS was used for calculation of
intensities in several linelists
\cite{jt283,jt336,jt365,jt365b,jt374,jt447} including the ExoMol linelist
 \cite{jt570}, where \ai\ and experimental energy levels were
merged. These linelists have been used in numerous astronomical studies,
the most outstanding of which was the detection of an atmosphere
around the super-Earth~55~Cancri~e~\cite{jt629}.  However, the
intensities obtained with this DMS deviate significantly from
the experimental ones \cite{95MaQuKl.HCN} for the ($0~0^0~1)$ 
fundamental band.
Harris \etal~\cite{jt283} already noted that the \ai\ \emph{P}
branch  line intensities are approximately 30~\%
stronger that the experimental ones and the second minimum of the
\emph{R} branch does not match the experimental minimum.

\section{Optimization procedure}
\label{sec.fit}
\subsection{General approach}

As we start from an \aipes\ with a high level of accuracy, our primary
goal is to tune it to accurately reproduce
the experimental data.  There are a number of methods of performing such
fits,
including, for example, morphing
\cite{99MeHuxx.methods,jt438}.  A general method of the optimization developed 
by Yurchenko \etal~\cite{03YuCaJe.PH3} was used.  Its
main idea is to fit a PES to empirical energies and \ai\ grid points
simultaneously to avoid nonphysical behavior of the fitting surface in
regions, which are poorly constrained by the observed data.  

%\blue{I have improved the notation below but since the coordinates
%for HCN and HNC are different, we may want to use different symbols for them.}

In the present work we construct different semi-empirical PESs for HCN
and HNC. In each case we start the fit from the \aipes\ for that well.
Performing a unified fit for both wells gave approximately the same
result ($\sigma \sim 0.3-0.4$ \cm) for both HCN and HNC.  This result
can be improved by treating each well separately but the unified PES
is important for studies of isomerization and states close to the
barrier.  Conversely, the need for accurate wavefunctions for
intensity predictions has been demonstrated in recent studies on other
molecules \cite{jt714,jt721}, which suggest that for best results
$\sigma$ needs to be below 0.05 \cm.  So far we have only been able to
 achieve this
level of accuracy by using polynomial functions localized in each well.

%\blue{This is actually a rather major step backwards
%and needs justifying. In particular it is difficult to see how this good
%for (a) tackling the isomerisation problem or (b) provides a good starting 
%point for a joint HCN/HNC line list.}
%\green{We do provide a good starting point for a joint HCN/HNC linelist because both these PES based on the same \ai\ data. If we were used the $V_{ai}$ PES for both wells we would have a poorer result because of the low sensitivity of the surface to small changes in the \ai\ data in pretty harmonic HCN well. I do not see this as a step backwards because here we are showing our methods and I think we will fit both wells with $V_{ai}$ after constructing new \aipes\, based on more \ai\ data (points).}

Considering first HNC, we added an additional surface to the
\aipes\ $V_{ai}$ expressed as an analytic polynomial:
\begin{equation}
\label{eq:form_HNC}
V^{\rm HNC}(r_1, r_2, \theta) = V_{ai}(r_1, r_2, \theta) + c_{ijk}{s_1}^i{s_2}^j{s_{\theta}}^k,
\end{equation}
$
s_1 = r_1 - r_{1e}^{\rm HNC},\\
s_1 = r_2 - r_{2e}^{\rm HNC},\\
s_{\theta} = \cos{\theta} - \cos{\theta_e^{\rm HNC}},\\
$
where $r_1$ is the CH distance, $r_2$ is the CN bond length,
$\theta$ is the H-C-N bond angle and the triple
$(r_{1e}^{\rm HNC},r_{2e}^{\rm HNC},\theta_e^{\rm HNC})$ corresponds to the equilibrium
configuration of HNC.

For the HCN well, which is more harmonic, we use a different functional representation. Namely, we fit the
same \ai\ data from I directly to a polynomial form:
\begin{equation}
\label{eq:poly_HCN}
V_{\rm ai}^{\rm HCN}(r_1, r_2, \theta) = b_{ijk}{s_1}^i{s_2}^j{s_{\theta}}^k.
\end{equation}
$
s_1 = r_1 - r_{1e}^{\rm HCN},\\
s_1 = r_2 - r_{2e}^{\rm HCN},\\
s_{\theta} = \cos{\theta} - \cos{\theta_e^{\rm HCN}},\\
$ where $r_1$ is the CH bond length, and  $r_2$ and $\theta$ are as defined
above, and $(r_{1e}^{\rm HCN},r_{2e}^{\rm HCN},\theta_e^{\rm HCN})$
corresponds to the equilibrium
configuration of HCN. 
This fit 
gives a standard deviation of about 1.5 \cm, which can be compared to 
the global of both wells performed in I, which gave  a standard deviation of about 2.6 \cm. 

$V_{\rm ai}^{\rm HCN}(r_1, r_2, \theta)$ provides the starting point
for a fit to the observed HCN energy levels, using the same method as
in the HNC case:
\begin{equation}
\label{eq:form_HCN}
V(r_1, r_2, \theta)^{\rm HCN} = V_{\rm ai}^{\rm HCN}(r_1, r_2, \theta) + d_{ijk}{s_1}^i{s_2}^j{s_{\theta}}^k.
\end{equation}

\begin{longtable}{p{0.4cm}p{0.4cm}p{0.4cm}r}
\caption{Coefficients of \ai\ points fit $b_{ijk}$ of the HCN polynomial function from eq.~\eqref{eq:poly_HCN}, powers of 10 in parenthesis.}
\label{tab:hcn_constai}\\
%\begin{tabular}
\hline\hline
$i$	&	$j$	&	$k$	&	$b_{ijk}$, \cm\ / \AA$^{i+j}$ \\
\hline\endfirsthead
$i$	&	$j$	&	$k$	&	$b_{ijk}$, \cm\ / \AA$^{i+j}$ \\
\hline\endhead
\multicolumn{4}{c}{\textit{Continued on next page...}} \\
\endfoot
\endlastfoot
0	&	0	&	0	&	-5.82538383470(1)	\\
1	&	0	&	0	&	3.86389321057(2)	\\
0	&	1	&	0	&	-5.73382774051(2)	\\
0	&	0	&	1	&	1.30358201578(4)	\\
2	&	0	&	0	&	1.56450000000(5)	\\
1	&	1	&	0	&	-1.02655390333(4)	\\
1	&	0	&	1	&	-7.34563934810(3)	\\
0	&	2	&	0	&	4.71857396141(5)	\\
0	&	1	&	1	&	-3.35986338937(4)	\\
0	&	0	&	2	&	2.53000000000(3)	\\
3	&	0	&	0	&	-2.99421924819(5)	\\
2	&	1	&	0	&	4.81418477567(3)	\\
2	&	0	&	1	&	-1.48980205956(3)	\\
1	&	2	&	0	&	1.03012191080(3)	\\
1	&	1	&	1	&	1.22487911460(4)	\\
1	&	0	&	2	&	-4.59100323190(3)	\\
0	&	3	&	0	&	-1.05384656549(6)	\\
0	&	2	&	1	&	7.74454197329(3)	\\
0	&	1	&	2	&	1.22794398106(3)	\\
0	&	0	&	3	&	1.60043377884(1)	\\
4	&	0	&	0	&	3.87806313524(5)	\\
3	&	1	&	0	&	-1.89291145139(4)	\\
3	&	0	&	1	&	4.57558177750(3)	\\
2	&	2	&	0	&	-1.59731251862(4)	\\
2	&	1	&	1	&	1.30743572477(4)	\\
2	&	0	&	2	&	6.80295483588(3)	\\
1	&	3	&	0	&	-4.30250739748(3)	\\
1	&	2	&	1	&	-3.34044719190(3)	\\
1	&	1	&	2	&	1.39132320354(4)	\\
1	&	0	&	3	&	-2.22723524934(3)	\\
0	&	4	&	0	&	1.44088983710(6)	\\
0	&	3	&	1	&	-1.14951539583(4)	\\
0	&	2	&	2	&	1.66693158444(4)	\\
0	&	1	&	3	&	1.55423061069(4)	\\
0	&	0	&	4	&	6.75636521337(2)	\\
5	&	0	&	0	&	-3.86359114243(5)	\\
4	&	1	&	0	&	3.94990741828(4)	\\
4	&	0	&	1	&	-8.27443936133(3)	\\
3	&	2	&	0	&	3.95163031431(4)	\\
3	&	1	&	1	&	4.64366715529(4)	\\
3	&	0	&	2	&	-6.21253840400(3)	\\
2	&	3	&	0	&	1.81395172040(4)	\\
2	&	2	&	1	&	-1.63118870226(3)	\\
2	&	1	&	2	&	-3.71978411806(4)	\\
2	&	0	&	3	&	-6.11942762545(3)	\\
1	&	4	&	0	&	-4.77484231577(3)	\\
1	&	3	&	1	&	1.65758045070(4)	\\
1	&	2	&	2	&	-3.30921689766(4)	\\
1	&	1	&	3	&	-1.74115345321(4)	\\
0	&	5	&	0	&	-1.36558717498(6)	\\
0	&	4	&	1	&	6.32814455194(4)	\\
0	&	3	&	2	&	-5.44011053142(3)	\\
0	&	2	&	3	&	-3.08293347406(4)	\\
0	&	1	&	4	&	-2.30801037828(4)	\\
0	&	0	&	5	&	-6.62318144340(2)	\\
6	&	0	&	0	&	2.06164152069(5)	\\
5	&	1	&	0	&	-2.72643462287(4)	\\
5	&	0	&	1	&	1.66406933717(4)	\\
4	&	2	&	0	&	-9.06418473036(4)	\\
4	&	1	&	1	&	-9.89426326122(4)	\\
4	&	0	&	2	&	-6.23124778630(3)	\\
3	&	3	&	0	&	2.20380959568(4)	\\
3	&	2	&	1	&	-9.35548681411(4)	\\
3	&	1	&	2	&	-1.61886146354(4)	\\
3	&	0	&	3	&	1.16982465743(4)	\\
2	&	4	&	0	&	0.00000000000(0)	\\
2	&	3	&	1	&	0.00000000000(0)	\\
2	&	2	&	2	&	-1.27099007662(4)	\\
2	&	1	&	3	&	2.31053677106(4)	\\
2	&	0	&	4	&	5.16397238424(3)	\\
1	&	5	&	0	&	0.00000000000(0)	\\
1	&	4	&	1	&	0.00000000000(0)	\\
1	&	3	&	2	&	0.00000000000(0)	\\
1	&	2	&	3	&	6.17471253018(4)	\\
1	&	1	&	4	&	1.20224840548(3)	\\
1	&	0	&	5	&	-1.57260975308(3)	\\
0	&	6	&	0	&	3.62568390524(5)	\\
0	&	1	&	5	&	1.40150909942(4)	\\
0	&	2	&	4	&	1.68845345324(4)	\\
0	&	3	&	3	&	1.55751022655(4)	\\
0	&	2	&	4	&	0.00000000000(0)	\\
0	&	1	&	5	&	0.00000000000(0)	\\
0	&	0	&	6	&	-4.05714839862(2)	\\
\hline\hline
%\end{tabular}
\end{longtable}

\begin{table}
\caption{Optimized coefficients $d_{ijk}$ of the HCN polynomial function from eq.~\eqref{eq:form_HCN}, powers of 10 in parenthesis.}
\label{tab:hcn_constfit}
\begin{tabular}{p{0.4cm}p{0.4cm}p{0.4cm}r}
\hline\hline
$i$	&	$j$	&	$k$	&	$d_{ijk}$, \cm\ / \AA$^{i+j}$ \\
\hline
1	&	0	&	0	&	-5.7427163997318(1)	\\
0	&	1	&	0	&	1.3345870332940(2)	\\
0	&	0	&	1	&	2.4975542880134(1)	\\
2	&	0	&	0	&	5.1497044820851(1)	\\
1	&	1	&	0	&	-1.4358942155508(2)	\\
1	&	0	&	1	&	-6.4634883126558(1)	\\
0	&	2	&	0	&	-1.6859489693547(2)	\\
0	&	1	&	1	&	-1.2665929510125(0)	\\
0	&	0	&	2	&	-4.9044301624532(1)	\\
3	&	0	&	0	&	2.2987168749403(2)	\\
0	&	3	&	0	&	-1.1715811331028(3)	\\
0	&	0	&	3	&	3.4727493153491(1)	\\
\hline\hline
\end{tabular}
\end{table}

\begin{table}
\caption{Optimized coefficients $c_{ijk}$ of the HNC polynomial function from eq.~\eqref{eq:form_HNC}, powers of 10 in parenthesis.}
\label{tab:hnc_const}
\begin{tabular}{p{0.4cm}p{0.4cm}p{0.4cm}r}
\hline\hline
$i$	&	$j$	&	$k$	&	$	c_{ijk}$, \cm\ / \AA$^{i+j}$ 	\\
\hline
0	&	0	&	0	&	$	-2.5228259785515(3)	$	\\
1	&	0	&	0	&	$	3.1876702203250(2)	$	\\
0	&	1	&	0	&	$	-6.6678874296762(3)	$	\\
0	&	0	&	1	&	$	-2.4842036191744(3)	$	\\
2	&	0	&	0	&	$	-1.5885178372949(3)	$	\\
1	&	1	&	0	&	$	2.0186290559122(3)	$	\\
1	&	0	&	1	&	$	6.6308401256739(3)	$	\\
0	&	2	&	0	&	$	2.1284923915787(4)	$	\\
0	&	1	&	1	&	$	-3.3052835812817(3)	$	\\
0	&	0	&	2	&	$	-1.7868381384170(4)	$	\\
3	&	0	&	0	&	$	2.1113127247948(3)	$	\\
0	&	3	&	0	&	$	-2.3906844780148(4)	$	\\
0	&	0	&	3	&	$	-1.8174121247121(5)	$	\\
4	&	0	&	0	&	$	3.1069836700747(3)	$	\\
0	&	0	&	4	&	$	-1.0227632425463(6)	$	\\
0	&	0	&	5	&	$	-1.6015153206752(6)	$	\\
\hline\hline
\end{tabular}
\end{table}

%We used corrections with different $(r_{1e},r_{2e},\theta_e)$ and set of constants for each of molecules (table \ref{tab:hnc_const}). 

\subsection{Nuclear motion calculations}
\label{sec.nucl}
The vibrational energy levels were calculated using the DVR3D program
suite \cite{jt338}.  The parameters used are presented in
Table~\ref{tab:dvr_param}; Morse-like oscillators were used for the
radial basis functions. We include almost all corrections used in
the \aipes: adiabatic and relativistic correction surfaces and a
nonadiabatic correction (by choosing atomic masses), to get
the best starting point for fitting.  More details about corrections
are given in I.

\begin{table}
\caption{Input parameters for DVR3DRJZ module of DVR3D \cite{jt338}; the Morse parameters are in atomic units, atomic masses are in Da.}
\begin{tabular}{lrl}
\label{tab:dvr_param} \\
\hline\hline
Parameter	&	Value	&	Description \\
\hline
NPNT1	&	40	&	No. of $r_1$ radial DVR points (Gauss-Laguerre) \\
NPNT2	&	40	&	No. of $r_2$ radial DVR points (Gauss-Laguerre) \\
NALF	&	50	&	No. of angular DVR points (Gauss-Laguerre) \\
NEVAL	&	950	&	No. of eigenvalues/eigenvectors required \\
MAX3D	&	5500	&	Dimension of final vibrational Hamiltonian \\
XMASS (H)	&	1.007825 &	Mass of hydrogen atom \\
XMASS (C)	&	12.000000 &	Mass of carbon atom \\
XMASS (N)	&	14.003074 &	Mass of nitrogen atom \\
$r_{1e}$	&	2.3	&	Morse parameter ($r_1$ radial basis function) \\
$D_{1e}$	&	0.1	&	Morse parameter ($r_1$ radial basis function) \\
$\omega_{1e}$	&	0.0105	&	Morse parameter ($r_1$ radial basis function) \\
$r_{2e}$	&	3.2	&	Morse parameter ($r_2$ radial basis function) \\
$D_{2e}$	&	0.1	&	Morse parameter ($r_2$ radial basis function) \\
$\omega_{2e}$	&	0.004	&	Morse parameter ($r_2$ radial basis function) \\
\hline\hline
\end{tabular}
\end{table}

\subsection{The HCN fit}
\label{sec.hcnfit}
We took $(r_{1e},r_{2e},\theta_e)$ = (1.066~\AA, 1.153~\AA,
$180.0^\circ)$ as the minimum in the HCN well from the \aipes. Mellau
\cite{11Mexxxx.HCN} reported experimental characterization of all 3822
eigenenergies of HCN up to 6880 \cm\ relative to the ground state
using high temperature hot gas emission spectroscopy. In nearly all
cases these empirical energy levels are accurate to better than 0.0004
\cm.  This dataset was used in this work.  The set of {\it ab initio}
and optimized polynomial coefficients from eq.~\eqref{eq:form_HCN} are
presented in Tables \ref{tab:hcn_constai} and \ref{tab:hcn_constfit},
respectively.

HCN vibrational ($J = 0$) energy levels are presented in table
\ref{tab:hcn_levels}.  A table of the calculated levels with $J =2$ 
is given in the supplementary material.
The standard deviation for the levels with $J = (0, 2,
  5, 9, 10)$ is $\sigma = 0.0373$ \cm.

\begin{longtable}{rrrcrc}
  \caption{ HCN energy levels computed using our recent \aipes\
    \cite{jt689} and the new PES, experimental data are taken from
    Mellau \etal~\cite{11Mexxxx.HCN}. All values are in \cm. 
   Experimental uncertainties for the levels are about 
     10$^{-4}$~\cm.}

\label{tab:hcn_levels} \\
\hline\hline
	($\nu_1$	&	$\nu_2$	&	$\nu_3$)	&	Obs.	&	$\text{Obs.} - \text{calc.}$	&	$\text{Obs.} - \text{calc.}$\\
	&		&		&		&	 \text{(\ai\ \cite{jt689})}	&	 \text{(this work)}	\\
\hline\endfirsthead
	($\nu_1$	&	$\nu_2$	&	$\nu_3$)	&	Obs.	&	$\text{Obs.} - \text{calc.}$	&	$\text{Obs.} - \text{calc.}$\\
	&		&		&		&	 \text{(\ai\ \cite{jt689})}	&	 \text{(this work)}	\\
\hline\endhead
\multicolumn{6}{c}{\textit{Continued on next page...}} \\
\endfoot
\endlastfoot

	0	&	2	&	0	&	1411.4135	&	0.18	&	-0.0357\\
	0	&	0	&	1	&	2096.8455	&	0.67	&	-0.0428\\
	0	&	4	&	0	&	2802.9587	&	0.71	&	-0.0306\\
	1	&	0	&	0	&	3311.4771	&	-1.25	&	-0.0052\\
	0	&	2	&	1	&	3502.1211	&	0.14	&	-0.0011\\
	0	&	0	&	2	&	4173.0709	&	1.09	&	-0.0712\\
	0	&	6	&	0	&	4174.6086	&	-0.08	&	0.0132\\
	1	&	2	&	0	&	4684.3100	&	-1.93	&	-0.0291\\
	0	&	4	&	1	&	4888.0393	&	1.13	&	-0.0337\\
	1	&	0	&	1	&	5393.6977	&	-0.68	&	-0.0132\\
	0	&	8	&	0	&	5525.8128	&	-1.54	&	0.0138\\
	0	&	2	&	2	&	5571.7343	&	0.23	&	0.0216\\
	1	&	4	&	0	&	6036.9601	&	-0.97	&	-0.0195\\
	0	&	0	&	3	&	6228.5983	&	1.34	&	-0.0640\\
	0	&	6	&	1	&	6254.4059	&	0.63	&	-0.0567\\
	2	&	0	&	0	&	6519.6105	&	-2.58	&	-0.0116\\
	1	&	2	&	1	&	6760.7051	&	-1.24	&	-0.0374\\
	0	&	10	&	0	&	6855.4431	&	-2.89	&	-0.0585\\
	0	&	4	&	2	&	6951.6830	&	1.58	&	-0.0531\\
	1	&	6	&	0	&	7369.4438	&	-	&	0.0418\\
	1	&	0	&	2	&	7455.4235	&	0.49	&	-0.0219\\
	0	&	8	&	1	&	7600.5358	&	-	&	-0.0478\\
\hline\hline
\end{longtable}

\subsection{The HNC fit}
\label{sec.hncfit}
We took $(r_{1e},r_{2e},\theta_e)$ = (2.187~\AA, 1.187~\AA,
$0.0^\circ)$ as the minimum of the HNC well.  The set of optimized
polynomial coefficients from eq.~\eqref{eq:form_HNC} are presented in
table \ref{tab:hnc_const}.

As for HCN, Mellau \cite{10Mexxxa.HNC,10Mexxxb.HNC,11Mexxxx.HNC}
performed extensive measurements of HNC energy levels and these data
were used for the fit.  HNC vibrational ($J = 0$) energy levels are
given in table \ref{tab:hnc_levels}, the standard deviation for the
levels with $J = (0, 2)$ is $\sigma = 0.37$ \cm.  A table of the
calculated levels with $J =2$ is given in the supplementary material.

\begin{longtable}{rrrcrc}
  \caption{ HNC vibrational energy levels computed with our recent \aipes\
      \cite{jt689} and the new PES, experimental data are taken from
    Mellau \cite{10Mexxxa.HNC,10Mexxxb.HNC,11Mexxxx.HNC}. All values
    are in \cm.}
\label{tab:hnc_levels} \\
%res.08.12.17	&	iter	&	4	&	
\hline\hline
	($\nu_1$	&	$\nu_2$	&	$\nu_3$)	&	Obs.	&	$\text{Obs.} - \text{calc.}$	&	$\text{Obs.} - \text{calc.}$\\
	&		&		&		&	 \text{(\ai\ \cite{jt689})}	&	 \text{(this work)}	\\
\hline\endfirsthead
	($\nu_1$	&	$\nu_2$	&	$\nu_3$)	&	Obs.	&	$\text{Obs.} - \text{calc.}$	&	$\text{Obs.} - \text{calc.}$\\
	&		&		&		&	 \text{(\ai\ \cite{jt689})}	&	 \text{(this work)}	\\
\hline\endhead
\multicolumn{6}{c}{\textit{Continued on next page...}} \\
\endfoot
\endlastfoot
\hline
0	&	2	&	0	&	926.5032	&	0.82	&	-0.4210	\\
0	&	4	&	0	&	1867.0587	&	4.61	&	-0.1969	\\
0	&	0	&	1	&	2023.8594	&	0.43	&	0.0390	\\
0	&	6	&	0	&	2809.2876	&	9.45	&	0.2325	\\
0	&	2	&	1	&	2934.8188	&	1.11	&	-0.4151	\\
1	&	0	&	0	&	3652.6566	&	-3.78	&	0.4372	\\
0	&	8	&	0	&	3743.7641	&	11.06	&	-0.2029	\\
0	&	4	&	1	&	3861.4285	&	4.29	&	0.1328	\\
0	&	0	&	2	&	4026.3981	&	1.96	&	-0.0580	\\
1	&	2	&	0	&	4534.4499	&	-2.89	&	-0.2187	\\
0	&	6	&	1	&	4790.8597	&	9.58	&	-0.6460	\\
0	&	2	&	2	&	4921.2449	&	2.20	&	-0.7661	\\
1	&	4	&	0	&	5428.9856	&	1.83	&	-0.3650	\\
1	&	0	&	1	&	5664.8527	&	-2.36	&	-0.5914	\\
0	&	4	&	2	&	5833.4283	&	7.31	&	-0.0032	\\
1	&	6	&	0	&	6322.7195	&	8.00	&	0.2206	\\
1	&	2	&	1	&	6532.4023	&	-1.12	&	-0.0580	\\
2	&	0	&	0	&	7171.4016	&	-0.58	&	0.0447	\\
1	&	8	&	0	&	7205.1561	&	9.06	&	-0.0174	\\
1	&	4	&	1	&	7413.9507	&	-	&	-0.0536	\\
\hline\hline
\end{longtable}

\section{Dipole moment surface construction}
\label{sec.dip_cons}

\subsection{Electronic structure computations}

As described elsewhere \cite{jt509}, the external field calculation
(ED) of the dipole moment points is preferable to the expectation
value method (XP).  The \ai\ surface computed for HCN in I gives, as a
byproduct, the expectation values of dipole moments. The corresponding PES
points are of high accuracy, so we thought that we could construct
a good DMS using these byproduct dipole moments. We fitted them to a
functional form similar to the one used below and the results were
very poor; some computed intensities differed by 50~\%\ from the
observed values and thus were much worse than the values given by the ED
DMS obtained 15 year ago using an aug-cc-pvQZ basis set and 
the default MOLPRO 
complete active space (CAS) \cite{jt273}. This demonstrates once more
the importance of using  the ED method to compute dipole
moments.

We employed the IC-MRCI all-electrons method for dipole moment
calculations as programmed in MOLPRO \cite{MOLPRO}.  In I, we gave a
detailed description of the choice of the CAS.  Selection of the CAS
is very important as it strongly influences accuracy of the
calculations.  Here, for the dipole moment calculations, we selected the
same CAS as  used in I for creating the \ai\ PES.  Using the 
 aug-cc-pCV6Z basis used in I proved to be prohibitively expensive for
the dipole moments calculations, as every dipole point requires 4
independent calculations of the energies in an external field. We
therefore used an aug-cc-pCV5Z basis instead to reduce computer time.
%Our estimates suggest that this choice should not degrade the accuracy of \red{the value} \blue{The value of what? The individual
%dipoles or the computed intensity? Do you have some 6Z ED dipoles we can compare with?} of more than \blue{??? 200? 1000? 0.4} percent.
We calculated 902 DMS points. In I the best result was achieved using the
Pople relaxed correction, so here  we decided to use the same MRCI correction 
for construction of the DMS. Also a relativistic correction, 
mass-velocity plus first-order Darwin term  (MVD1) as produced by MOLPRO was added. Beyond Born-Oppenheimer effects have been shown to be fairly
small \cite{09HoVaEd.DMS} and are not considered here.

\subsection{Fitting the DMS}
A polynomial form similar to eq.~\eqref{eq:poly_HCN} was used to represent 
the DMS:
\begin{equation}
\label{eq:poly_dmsx}
\mu_x(r_1, r_2, \theta) = t_{ijk}{s_1}^i{s_2}^j{s_{\theta}}^k,
\end{equation}
\begin{equation}
\label{eq:poly_dmsz}
\mu_z(r_1, r_2, \theta) = u_{ijk}{s_1}^i{s_2}^j{s_{\theta}}^k.
\end{equation}
A total of 82 parameters (for each component) were determined from the
fit to 902 dipole moment points that resulted in a standard deviation
of $2.56 \times 10^{-6}$ a.u. from the fitted surface for the $x$
dipole component, and  $5.16 \times 10^{-6}$
a.u. for the $z$ component.

\section{Intensity analysis}
\label{sec.intens}

We computed transition intensities for HCN transitions up to $J = 10$
for wavenumbers up to 7~200 \cm\ using the PESs and DMSs described
above. The transition intensities of each of the fundamentals has recently been measured
experimentally and they are detailed below.

\subsection{PES part}
Table II in the Supplementary Material shows the differences in intensities
caused by the improvement in accuracy of wavefunctions for
lines with intensities higher than $10^{-23}$ cm/molecule. These
transitions were grouped by differences in percentages and the results
of this are presented in the table \ref{tab:summ_intens}.  It is
apparent that transition intensities obtained with the same DMS may
vary significantly with wavefunctions calculated using
the \aipes\ from I or the new fitted one.
\begin{table}
\caption{Comparison of HCN intensities computed using potential surfaces by 
van Mourik \etal~\cite{jt273}, Makhnev~\etal \cite{jt689} (I) and using the new 
fitted PES. The comparison is made for 550 transitions between the states 
with $J \le 5$. The DMS used in the calculations was taken from \cite{jt273}. 
The percentage differences are computed as $(I_{1} / I_{2} - 1) \cdot 100$~\%.}
\label{tab:summ_intens}
\begin{tabular}{lcc}
\hline\hline
	&\multicolumn{2}{c}{$N_{\textrm{lines}}$ between PESs from:} \\
Intensity difference & \cite{jt273} -- \cite{jt689} & \cite{jt689} -- this work\\
	&	(old -- new \ai)	&	(new \ai\ -- fitted)	\\
\hline
$< 1~\%$ & 255 & 487 \\
$1-5~\%$ & 232 & 35 \\
$5-10~\%$ & 22 & 18 \\
$>10~\%$ & 41 & 10 \\
%\multicolumn{3}{r}{out of 550 lines}	\\
\hline\hline
\end{tabular}
\end{table}

\subsection{DMS part}
We compared our predicted intensities with available experimental data to 
verify the quality of our surfaces.  The mean differences between the experimental and
calculated intensities are shown in table \ref{tab:summ_trans}. For
all fundamental transitions (bold) the new DMS  better reproduces
the measurements. Accuracy approaching 1~\%\  is
achieved for two of the fundamentals, the H -- CN stretch and the bend.
These predictions lie
within the quoted experimental uncertainties. 
The behavior of the
CN-stretch is described in detail below.

The region about the HCN well has previously been extensively studied
theoretically; the improved accuracy of our calculated intensities for the
fundamental bands lies, mostly, in two factors: use of an
increased density of dipole moment points and a larger basis set.  The
influence of both of these factors for each group of transitions is
considered separately.

We also computed and fitted about 300 dipole points near the HCN
minimum with the same level of theory as that  used by van Mourik
\etal~\cite{jt273} and fitted these using eqs.  \eqref{eq:poly_dmsx}
and \eqref{eq:poly_dmsz}. Intensities obtained using these points
help to quantify the dependence of the results on the number of \ai\
points used in the fit, since only 115 points were used to determine
the DMS by van Mourik \etal~\cite{jt273}.
Below we call this new DMS the ``new local'' one, the DMS from van Mourik \etal
\cite{jt273} is described as the ``old global'' one, and our improved
DMS (described in section \ref{sec.dip_cons}) is simply called the
'`new'' one.

\begin{table}
\caption{Standard deviation of difference ($\varepsilon = I_{\text{calc}} / I_{\text{exp}} - 1,\,\%$) 
between the calculated and observed HCN transition intensities.% The experimental sources will be described in the text.
}
\label{tab:summ_trans}
\begin{tabular}{cccccl}
\hline\hline
Transition	&	DMS \cite{jt273}	&	DMS (this work)	& $N_{\text{lines}}$	&	Exp. uncert., \%	&	Exp. source	\\
\hline
$\nu_{1}$	&	\textbf{1.60}	&	\textbf{1.24}	&	40	&	2-3		&	\cite{03DevyNU1}	\\
$\nu_{1}+\nu_{2}^{1}-\nu_{2}^{1}$, (e/f)	&	5.21/5.05	&	5.23/5.07	&	35/41	&	4-5	&	\cite{03DevyNU1}	\\
$\nu_{2}^{1}$	&	\textbf{2.40}	&	\textbf{1.48}	&	50	&	2-3	&	\cite{08Smith.low}	\\
$2\nu_{2}^{0}$	&	2.31	&	0.98	&	40	&	2-3	&	\cite{04Devy2NU2}	\\
$2\nu_{2}^{0}-\nu_{1}^{1}$	&	5.89	&	5.80	&	36	&	4-6	&	\cite{08Smith.low}	\\
$2\nu_{2}^{2}-\nu_{1}^{1}$	&	6.95	&	6.59	&	45	&	4-10	&	\cite{08Smith.low}	\\
$3\nu_{2}^{1}-\nu_{2}^{1}$	, (e/f)	&	3.99/4.87	&	3.98/4.68	&	33/35	&	4-6	&	\cite{04Devy2NU2}	\\
$2\nu_{2}^{2}$	&	6.30	&	6.74	&	37	&	[5]$^{a}$	&	\cite{97MaQuKl.HCN}	\\
$3\nu_{2}^{3}$	&	12.32	&	15.39	&	14	&	[5]$^{a}$	&	\cite{97MaQuKl.HCN}	\\
$2\nu_{2}^{2}+\nu_{3}$	&	19.54	&	14.10	&	30	&	[5]$^{a}$	&	\cite{97MaQuKl.HCN}	\\
$\nu_{1}+2\nu_{2}^{2}$	&	48.99	&	23.95	&	19	&	[5]$^{a}$	&	\cite{97MaQuKl.HCN}	\\
$(\nu_{3})^b$	&	\textbf{34.42}	&	\textbf{16.58}	&	19	&	5	&	\cite{95MaQuKl.HCN}	\\
\hline\hline

\multicolumn{6}{l}{$^{a}$ Maki \etal~\cite{97MaQuKl.HCN} do not give experimental uncertainties. The value $5 \%$ is based  on } \\
\multicolumn{6}{l}{	previous work of Maki group \cite{95MaQuKl.HCN} with the same experimental setup.}\\
\multicolumn{6}{l}{$^{b}$ P-branch only.}\\
\end{tabular}
\end{table}

\subsubsection{HCN $\nu_{1}$, $\nu_{1}+\nu_{2}^{1}-\nu_{2}^{1}$ transitions (stretching of H--C)}

Experimental data were taken from Devi \etal~\cite{03DevyNU1}

\begin{figure}[]
\vspace{-3.25cm}

\center{\includegraphics[scale=0.9,trim= 50 0 0 0]{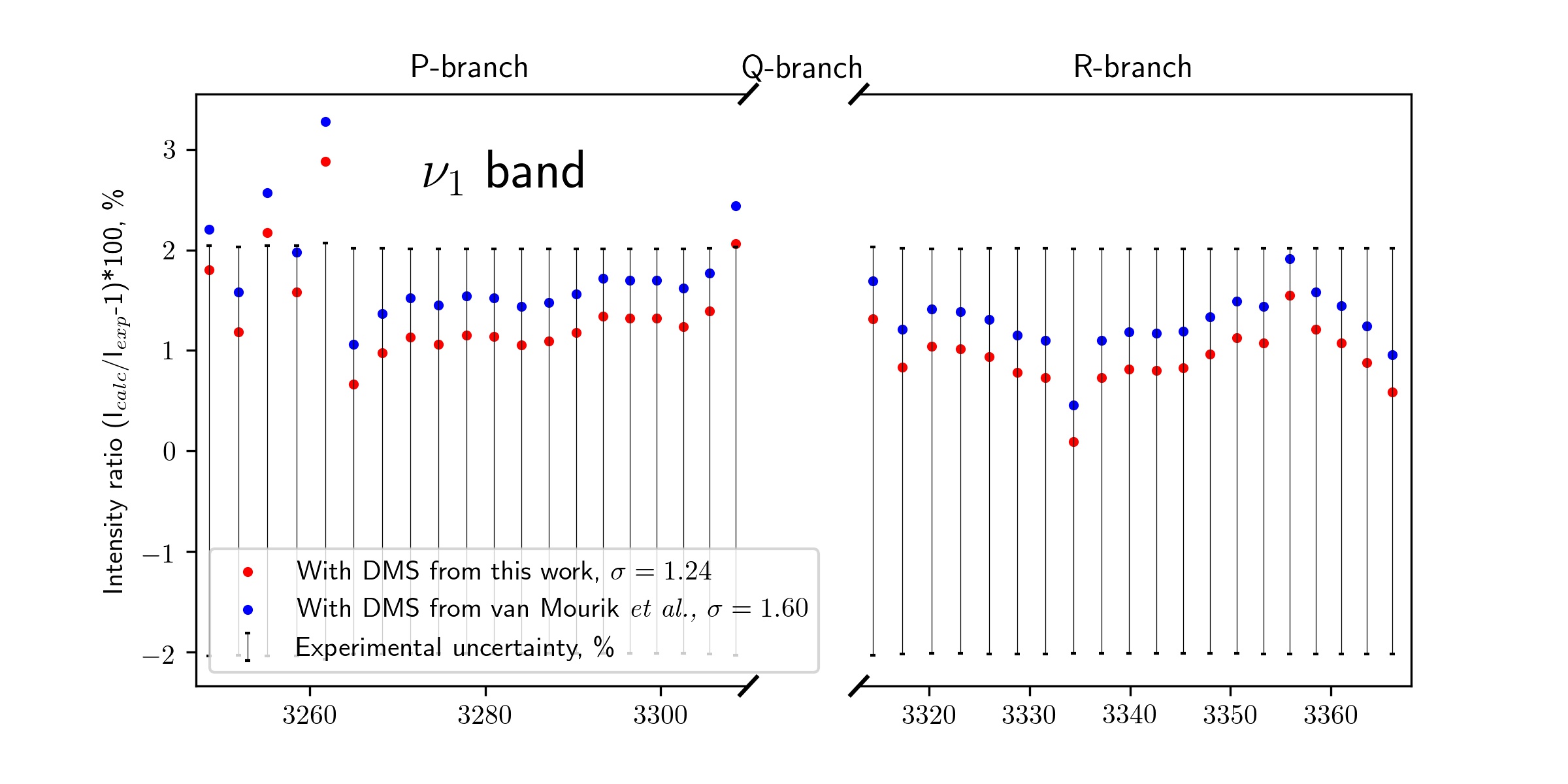}}
%\caption{Comparison of line intensitions for the ($1000-0000$) band.}
%\label{fig:nu1}
%\end{figure}

\vspace{-1.5cm}

%\begin{figure}[]
\center{\includegraphics[scale=0.9,trim= 50 0 0 0]{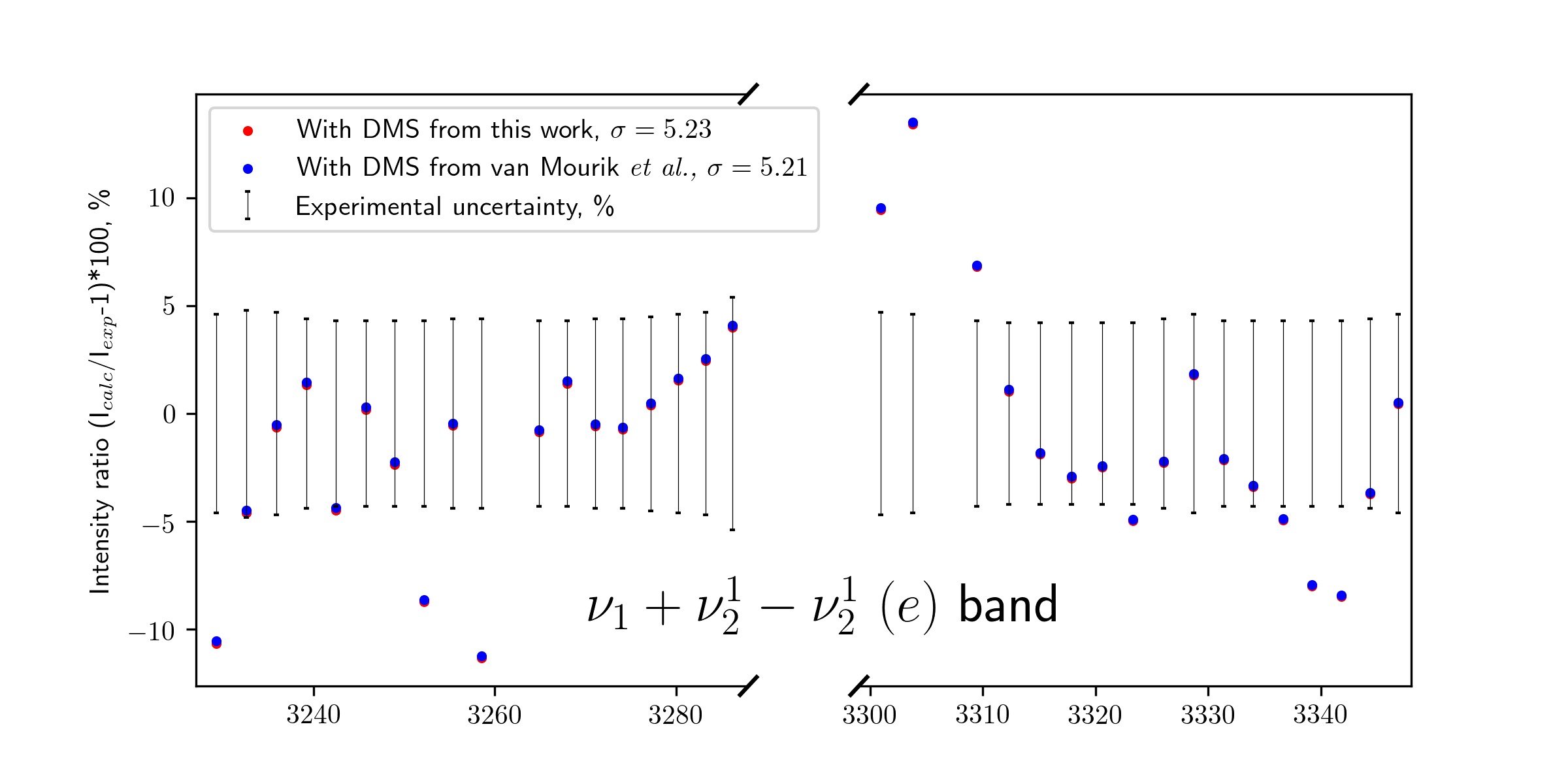}}
%\caption{Comparison of line intensitions for the (($1110-0110$)(e)  band.}
%\label{fig:enu1nu21}
%\end{figure}

\vspace{-1.5cm}

%\begin{figure}[]
\center{\includegraphics[scale=0.9,trim= 50 0 0 0]{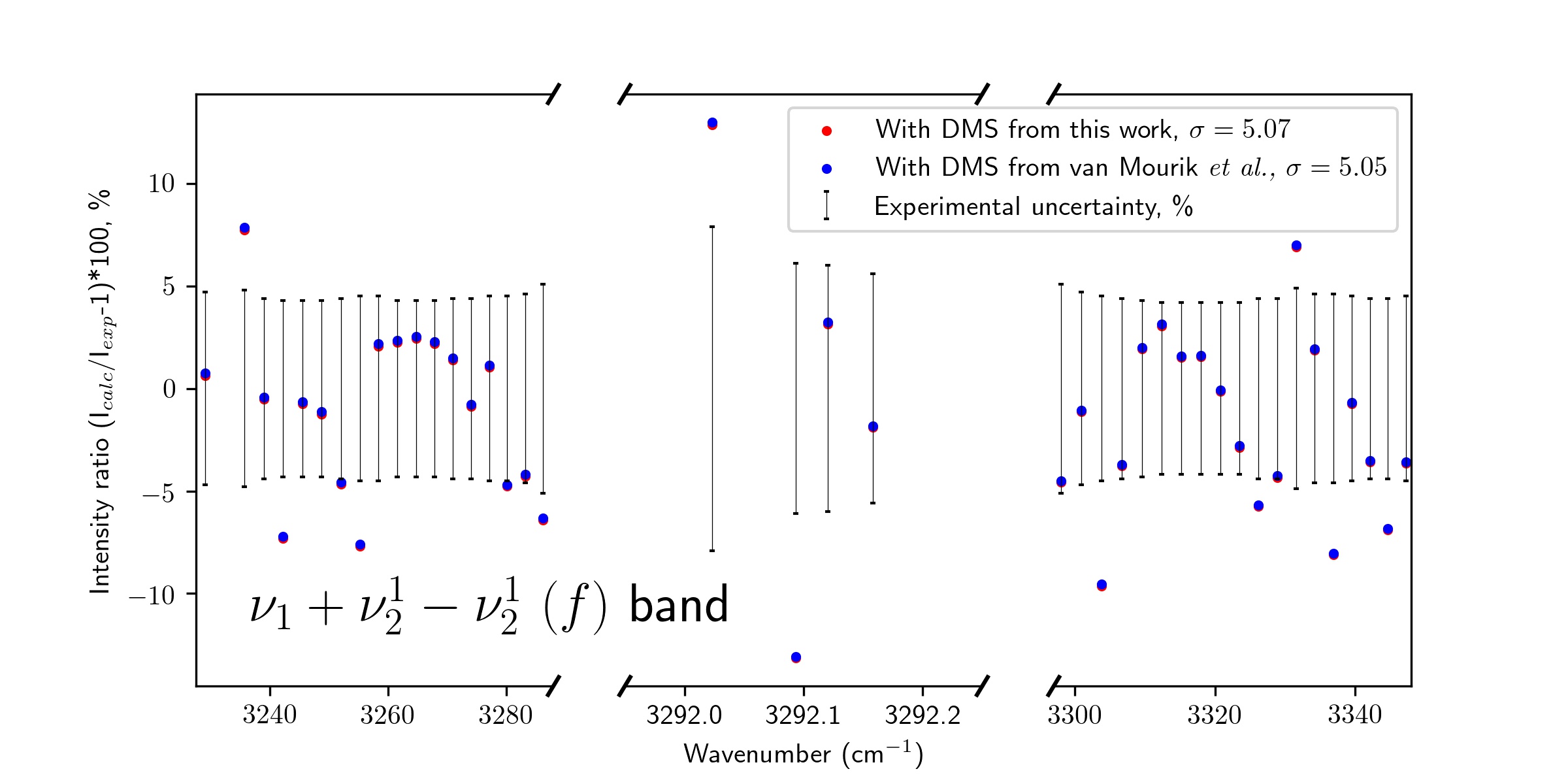}}
\caption{Comparison of HCN line intensities for the $1000-0000$ (top),  $1110-0110$(e) (middle) and $1110-0110$(f) (bottom)  bands.}
\label{fig:nu1}
\end{figure}

The H--C coordinate comprises the strongest type of
covalent chemical bond -- a $\sigma$ bond. But since this bond is
single, its properties can be described relatively easily, and a
satisfactory level of accuracy for the H--C vibrational
mode had been already achieved some time
ago. Figure \ref{fig:nu1} shows an improvement between the calculation
performed with the ``old global'' DMS \cite{jt273} (red) and the one
with the new DMS (blue) for a $\nu_{1}$ transition. The standard deviation of the intensities for this 
band improves from 1.60 to 1.24~\%. For the hot 
band $\nu_{1}+\nu_{2}^1-\nu_{2}^1$ (figure \ref{fig:nu1}) there are no 
significant changes but there is a huge difference between the corresponding $e/f$ 
transitions. 

For $\nu_1$ transitions the difference between the ``old global'' and
the recalculated ``new local'' DMSs is much smaller than 1~\%, so
there is only a weak dependence on the number of points here, and the
main reason for the improved accuracy of calculations achieved with
the new DMS is a better basis set used in
the \ai\ calculations.

\subsubsection{HCN $\nu_{2}^{1}$, $2\nu_{2}^{0}$, $2\nu_{2}^{0}-\nu_{2}^{1}$, 
$2\nu_{2}^{2}-\nu_{1}^{1}$, $3\nu_{2}^{1}-\nu_{2}^{1}$ transitions (bending)}

Experimental data were taken from Devi \etal~\cite{04Devy2NU2} 
and Smith \etal~\cite{08Smith.low}

\begin{figure}[]
\vspace{-3cm}
\center{\includegraphics[scale=1.0,trim= 50 0 0 0]{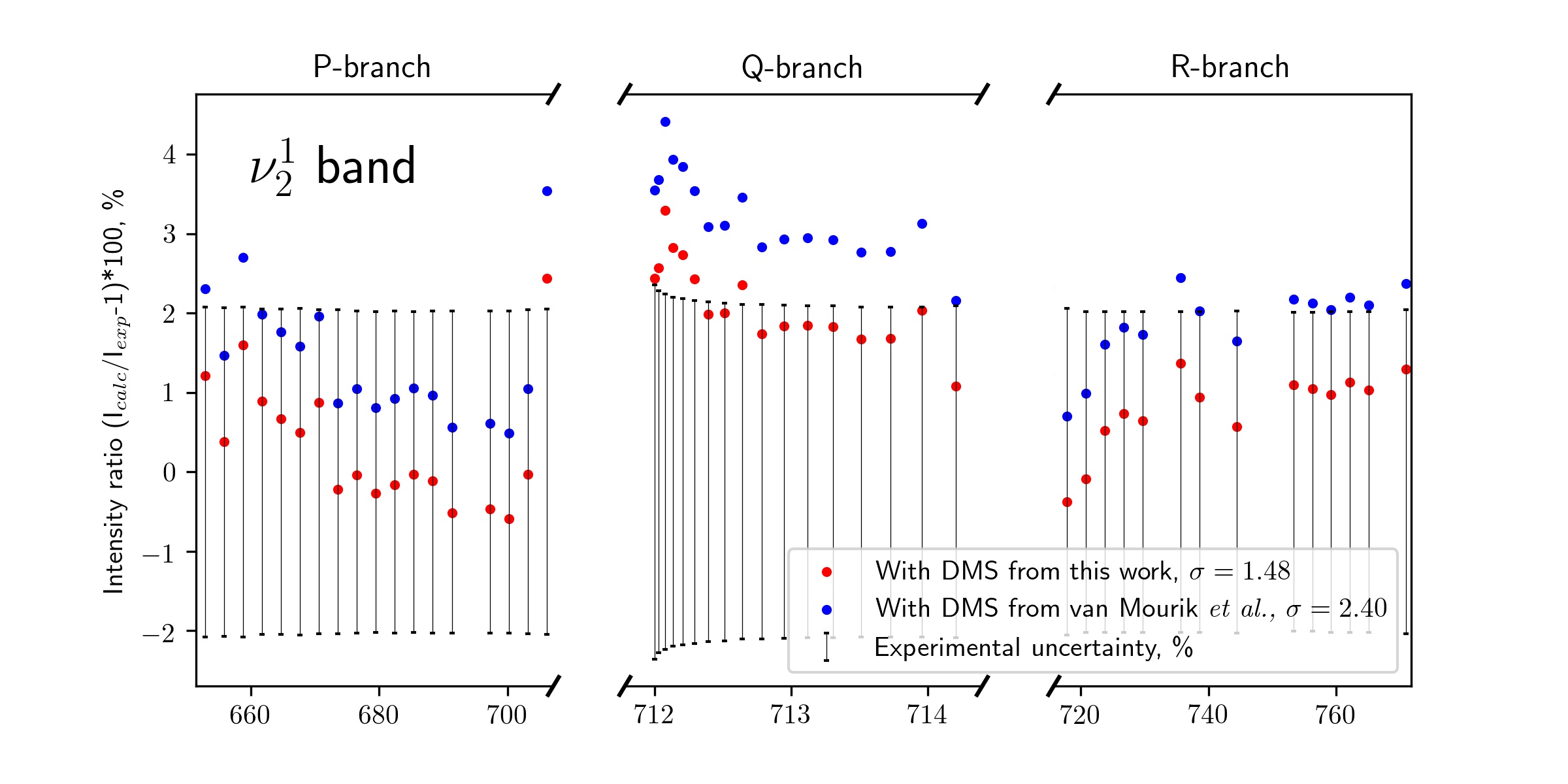}}
%\caption{Comparison of line intensitions for the ($0110-0000$)  band.}
%\label{fig:nu21}
%\end{figure}

\vspace{-1cm}

%^\begin{figure}[]
\center{\includegraphics[scale=1.0,trim= 50 0 0 0]{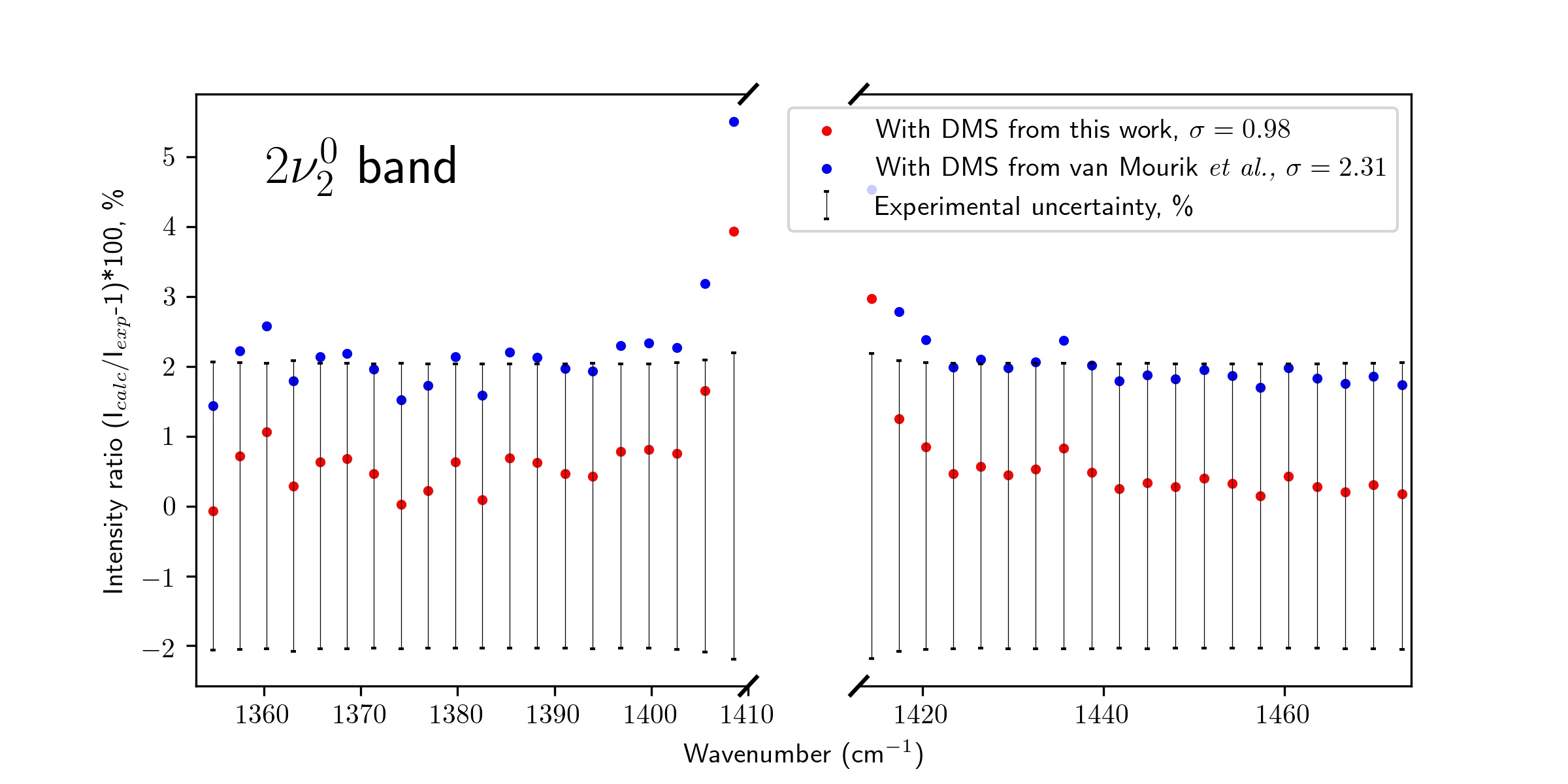}}
\caption{Comparison of HCN line intensities for the  $0110-0000$ (upper) and $0200-0000$ (lower)  bands.}
\label{fig:nu2}
\end{figure}

\begin{figure}[]
\vspace{-3cm}
\center{\includegraphics[scale=1.0,trim= 50 0 0 0]{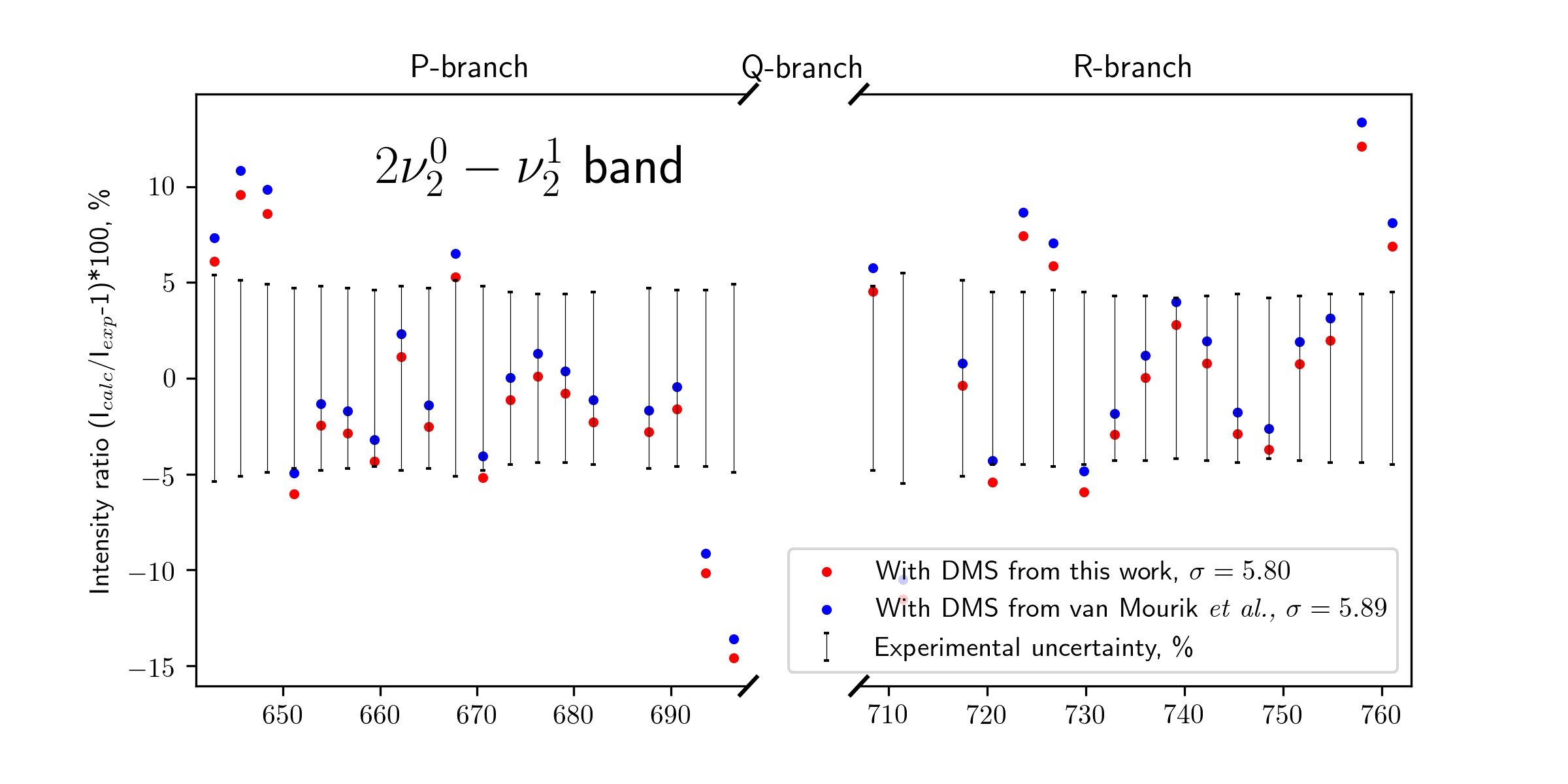}}
%\caption{Comparison of line intensitions for the ($0200-0110$)  band.}
%\label{fig:2nu20-nu21}
%\end{figure}

\vspace{-1cm}

%\begin{figure}[]
\center{\includegraphics[scale=1.0,trim= 50 0 0 0]{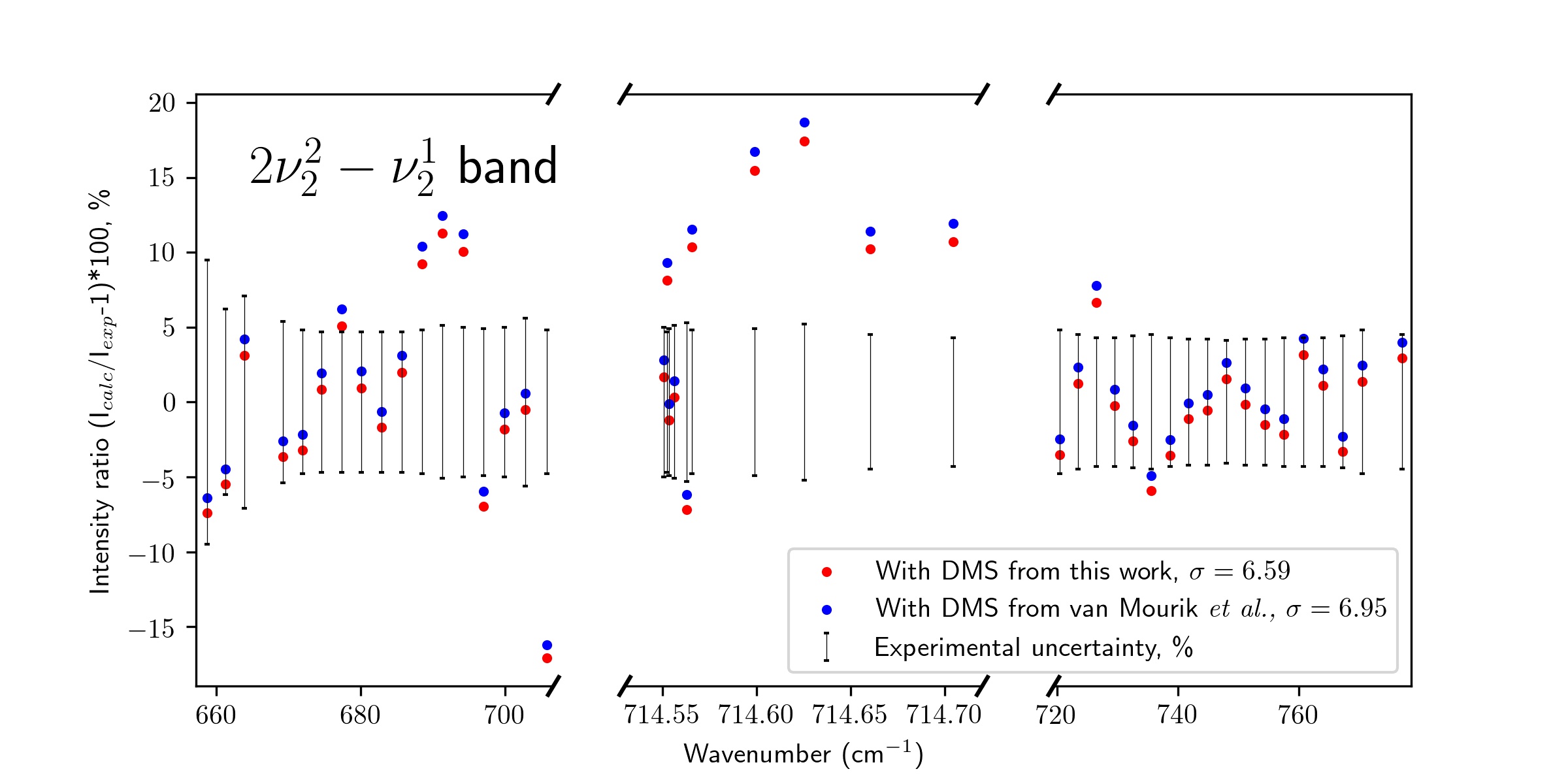}}
\caption{(Comparison of HCN line intensities for the $0200-0110$ (upper) and  $0220-0110$ (lower)  bands.}
\label{fig:nu2hot}
\end{figure}

\begin{figure}[]
\vspace{-3cm}
\center{\includegraphics[scale=1.0,trim= 50 0 0 0]{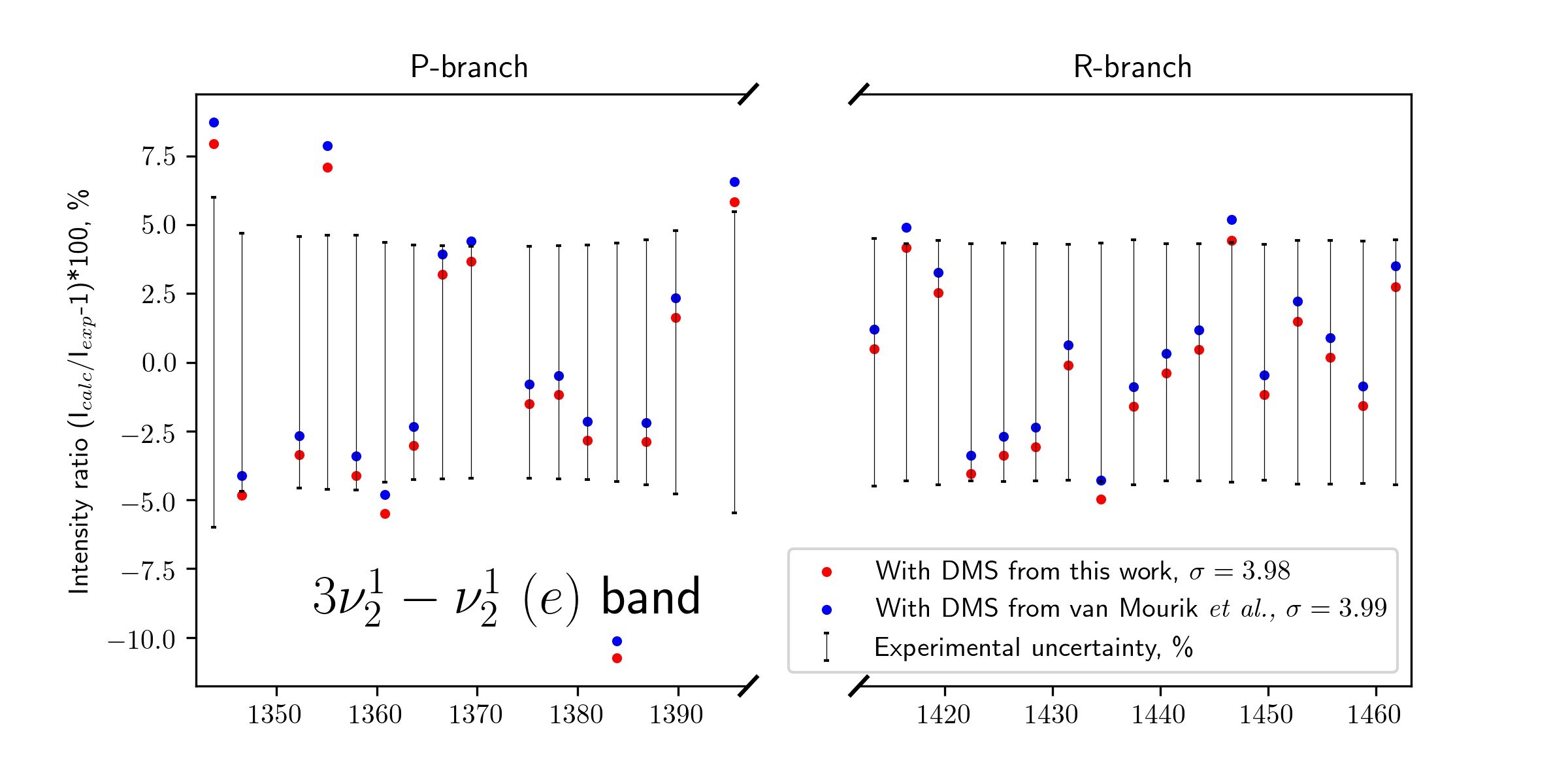}}
%\caption{Comparison of line intensitions for the ($0310-0110$)(e)  band.}
%\label{fig:e3nu21-nu21}
%\end{figure}

\vspace{-1cm}

%\begin{figure}[]
\center{\includegraphics[scale=1.0,trim= 50 0 0 0]{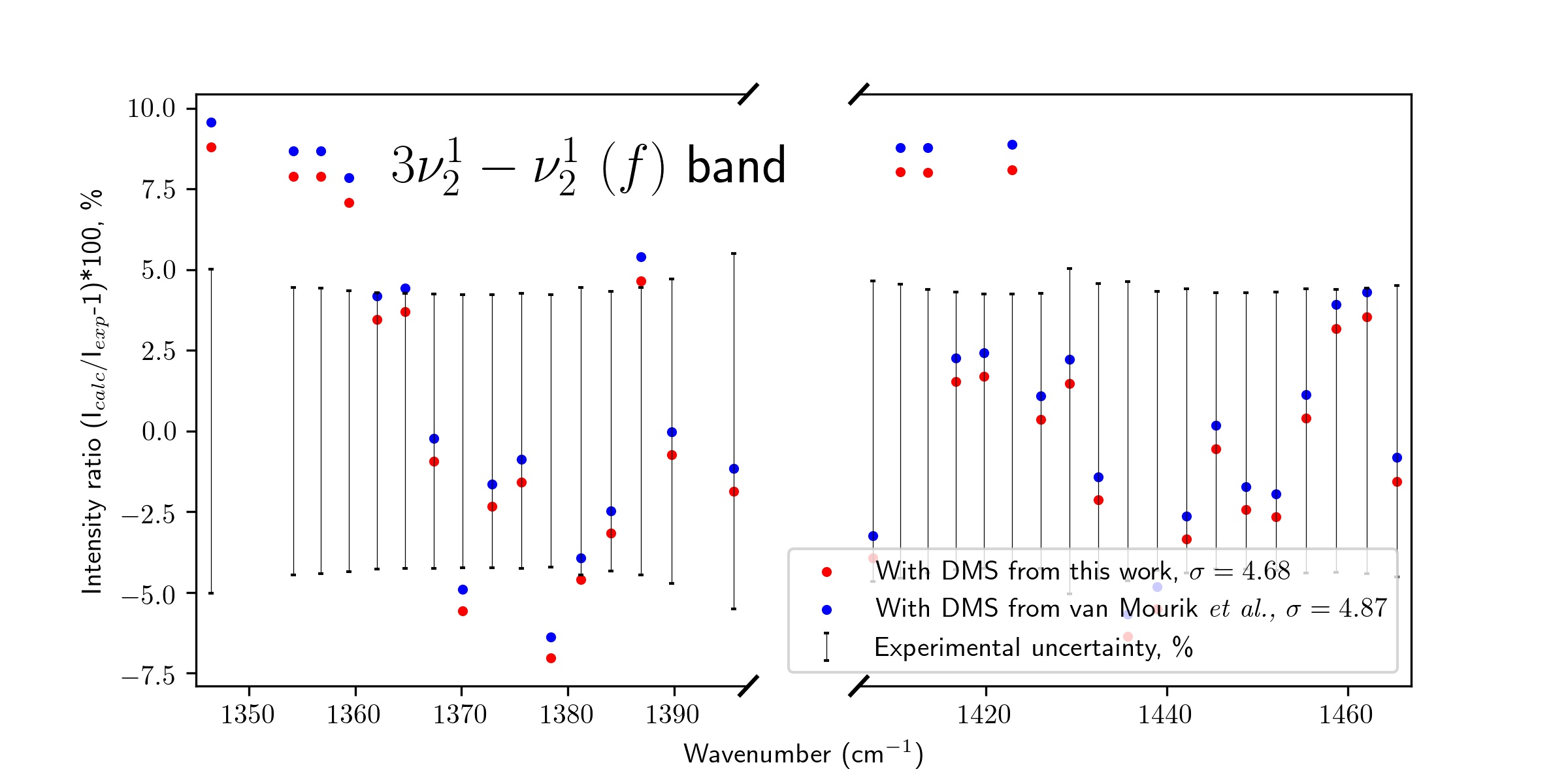}}
\caption{Comparison of HCN line intensities for the $0310-0110$(e) (upper) and $0310-0110$(f)  (lower) bands.}
\label{fig:3nu21-nu21}
\end{figure}

In contrast to the H--C stretch, bending motions probe a weaker
$\pi$-bond and harder to analyze with the same level of accuracy. But
for the $\nu_{2}^1$ and $2\nu_{2}^0$ transitions (see figure
\ref{fig:nu2}) the intensity differences are reduced from 2.40
and 2.31~\% to 1.48 and 0.98~\%, respectively. The accuracy of ``hot''
transitions ($2\nu_{2}^0-\nu_{2}^1$ -- figure \ref{fig:nu2hot},
$2\nu_{2}^{2}-\nu_{1}^{1}$ -- figure \ref{fig:nu2hot},
$3\nu_{2}^{1}-\nu_{2}^{1}$ -- figures \ref{fig:3nu21-nu21}) intensity
calculations is almost unimproved.

For this group of transitions the dependence of the computed
intensities on the number of dipole points used increases to 8-12~\%.
The resulting percent and sub-percent accuracy we obtain relies on a
combination of both factors: increased number of points and improved
level of theory.
%%%%%%%%%%%%%%%%%%%%%%%%%%%%%%%%%%%%%%%%
%%%%%%%%%%%%%%%%%%%%%%%%%%%%%%%%%%%%%%%%%%%
%%%%%%%%%%%%%%%%%%%%%%%%%%%%%%%%%%%%%%%%%

\subsubsection{HCN $\nu_{3}$ transitions (stretching C$\equiv$N)}

The atoms in the CN-stretch are connected by a triple bond
($\sigma+2\pi$) involving six electrons, which is much harder to
handle theoretically.  A comparative study of this CN-mode was
performed by Maki \etal~\cite{95MaQuKl.HCN}, who connected the
existence of a second minimum in the intensity of the R branch
transitions near R(7) for the main isotopologue with strong
vibration-rotation interaction for this mode.  The minimum (or the
gap) shifts between isotopologues and can also be detected in
the (01$^1$1--01$^1$0) hot band. Intensities computed for this stretch mode
with the ``old global'' DMS by Harris \etal~\cite{jt283} are of the
same order of magnitude.

Figure \ref{fig:nu3-int} shows the calculated intensities of the P-branch 
from the (0001--0000) band with the ``old global'' (blue), 
the ``new local'' (black) and the new (red) DMSs. 
The experimental data are taken from Maki \etal~\cite{95MaQuKl.HCN} (green). 
Here increasing the number of calculated points changes the calculated intensities by
about 20~\%. 

\begin{figure}[]
\vspace{-3cm}
\center{\includegraphics[scale=1.0,trim= 50 0 0 0]{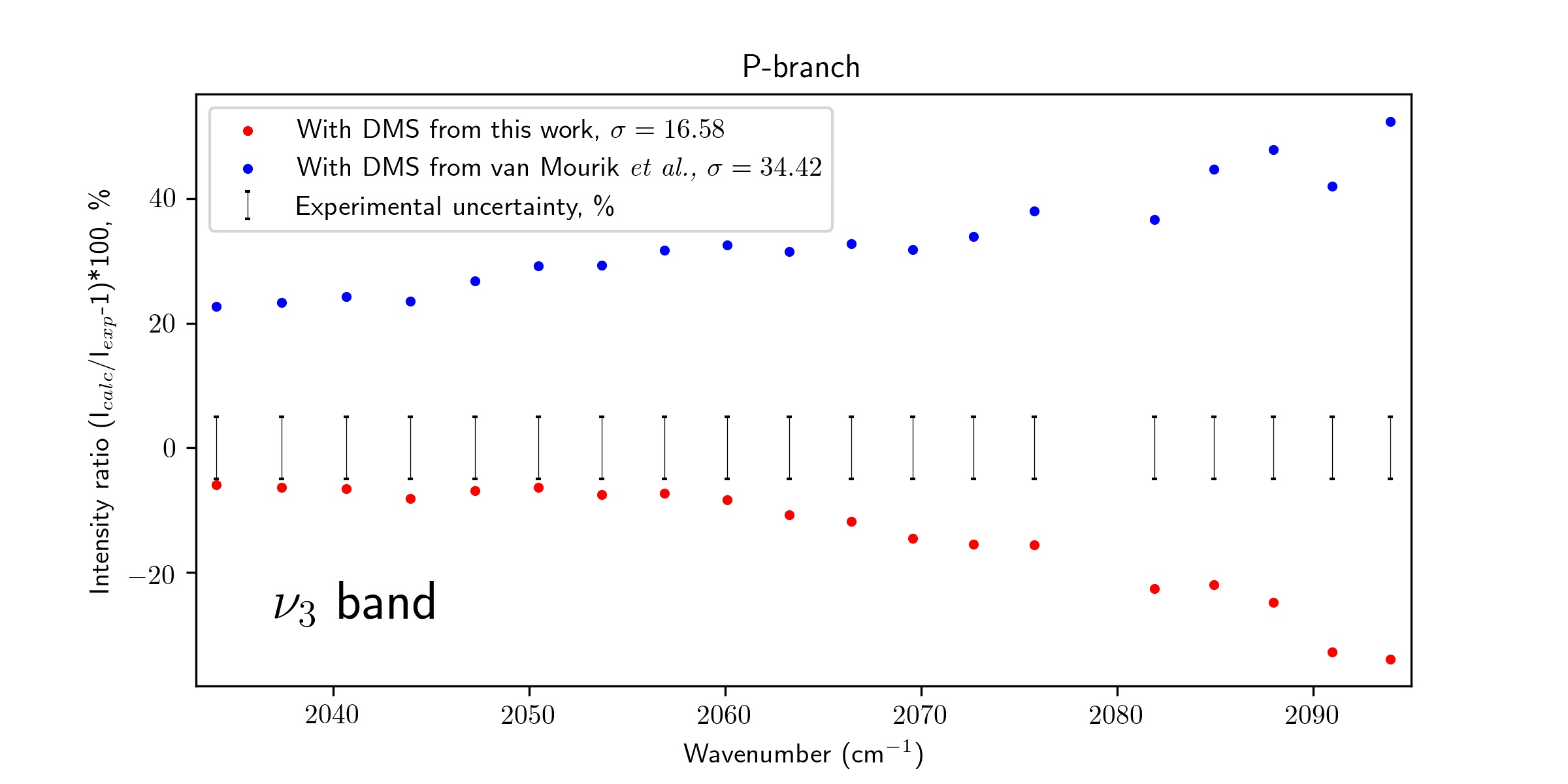}}
\caption{Comparison of line intensities for the ($0001-0000$) transition, P-branch.}
\label{fig:nu3}
\end{figure}

\begin{figure}[]
\center{\includegraphics[scale=1.0,trim= 50 0 0 0]{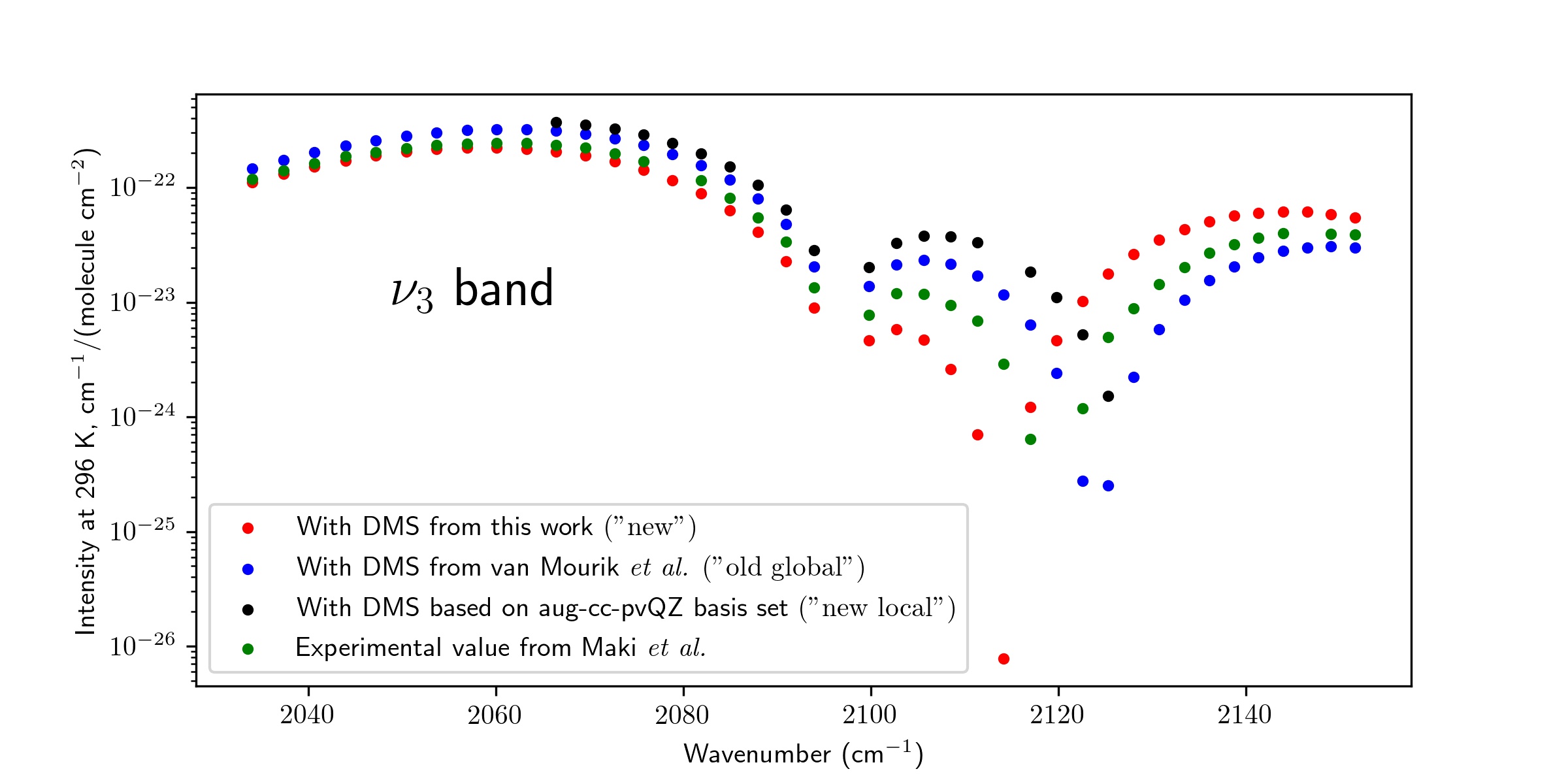}}
\caption{Comparison of line intensities for the ($0001-0000$)  band.}
\label{fig:nu3-int}
\end{figure}

Intensities calculated with the new DMS give improved agreement: the
deviation from the experimental data are 21.5~\% in comparison with
36.2~\% for the ``old global'' DMS (see figure \ref{fig:nu3}). It is
likely that further improvement will require both higher level calculations
and a denser grid.

\subsubsection{HCN transitions with $\Delta(l) > 1$: $2\nu_{2}^{2}$, $3\nu_{2}^{3}$ 
(bending), $2\nu_{2}^{2}+\nu_{3}$ (bending + stretching C$\equiv$N), $\nu_{1}+2\nu_{2}^{2}$ (bending + stretching C--H)}

Experimental data were taken from Maki \etal~\cite{97MaQuKl.HCN}

\begin{figure}[]
\vspace{-3cm}
\center{\includegraphics[scale=1.0,trim= 50 0 0 0]{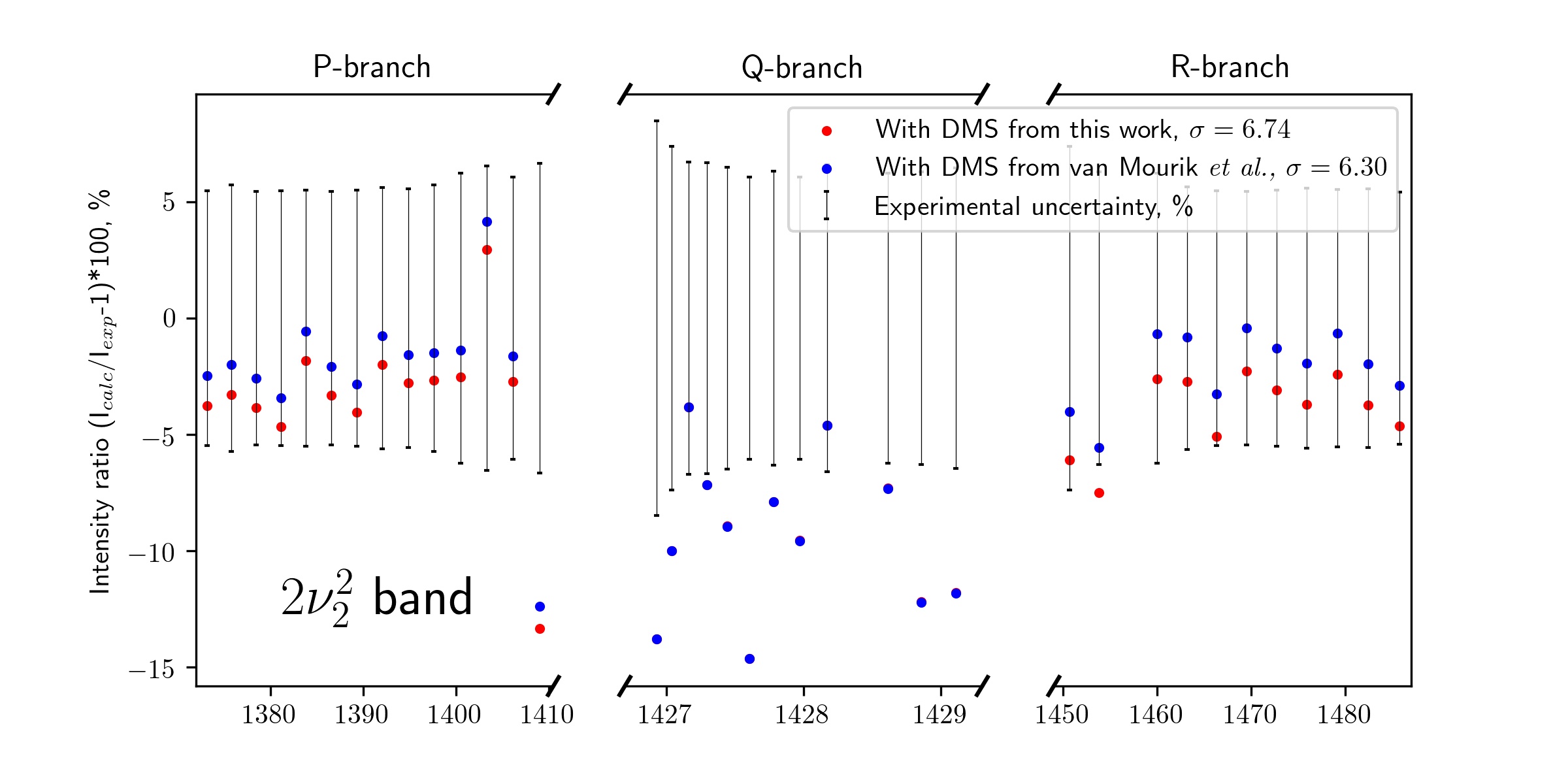}}
%\caption{Comparison of line intensitions for the ($0220-0000$)  band.}
%\label{fig:2nu22}
%\end{figure}

\vspace{-1cm}

%\begin{figure}[]
\center{\includegraphics[scale=1.0,trim= 50 0 0 0]{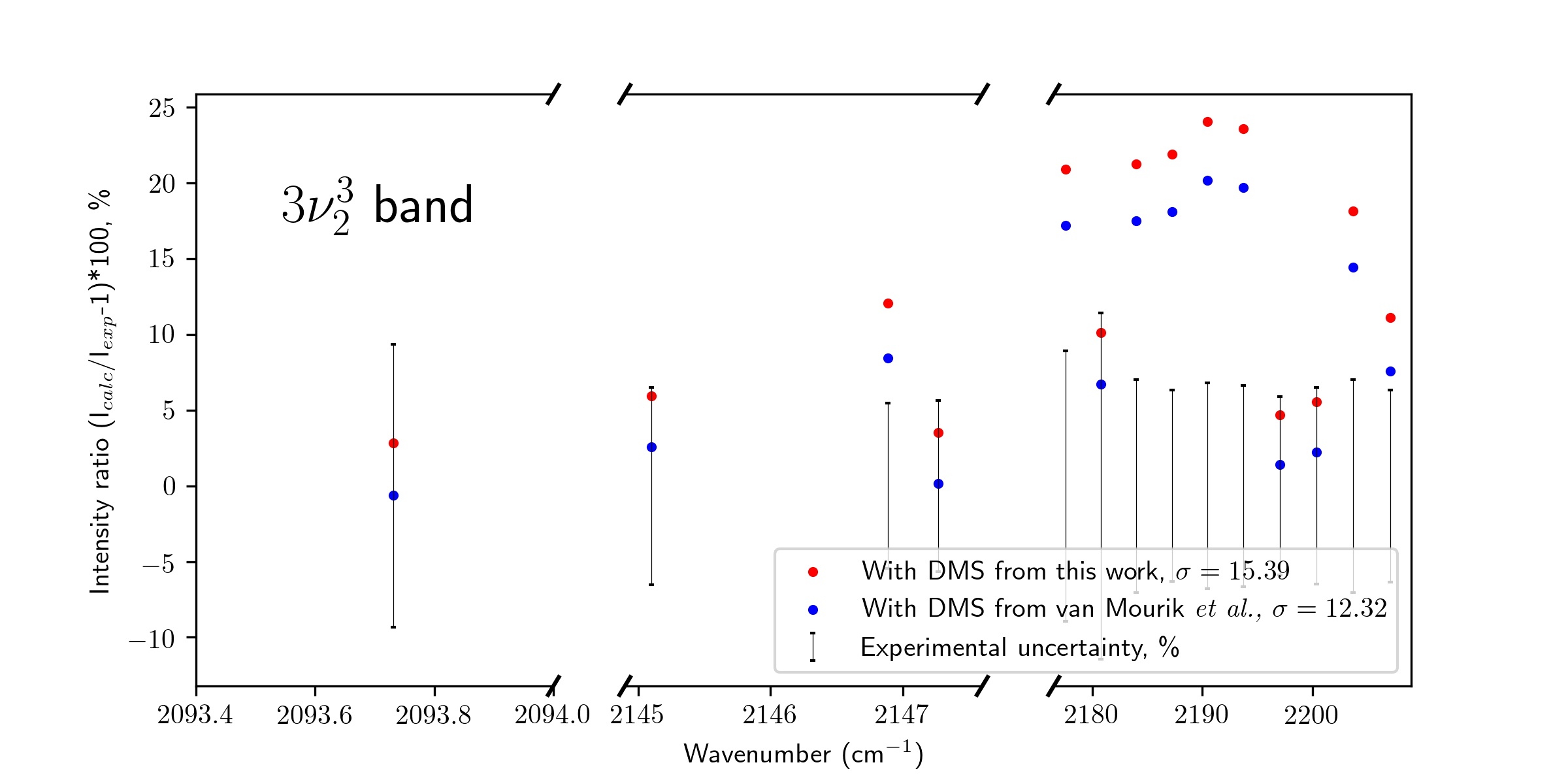}}
\caption{Comparison of line intensities for the ($0220-0000$) (upper) and ($0330-0000$) (lower) bands.}
\label{fig:3nu23}
\end{figure}

\begin{figure}[]
\vspace{-3cm}
\center{\includegraphics[scale=1.0,trim= 50 0 0 0]{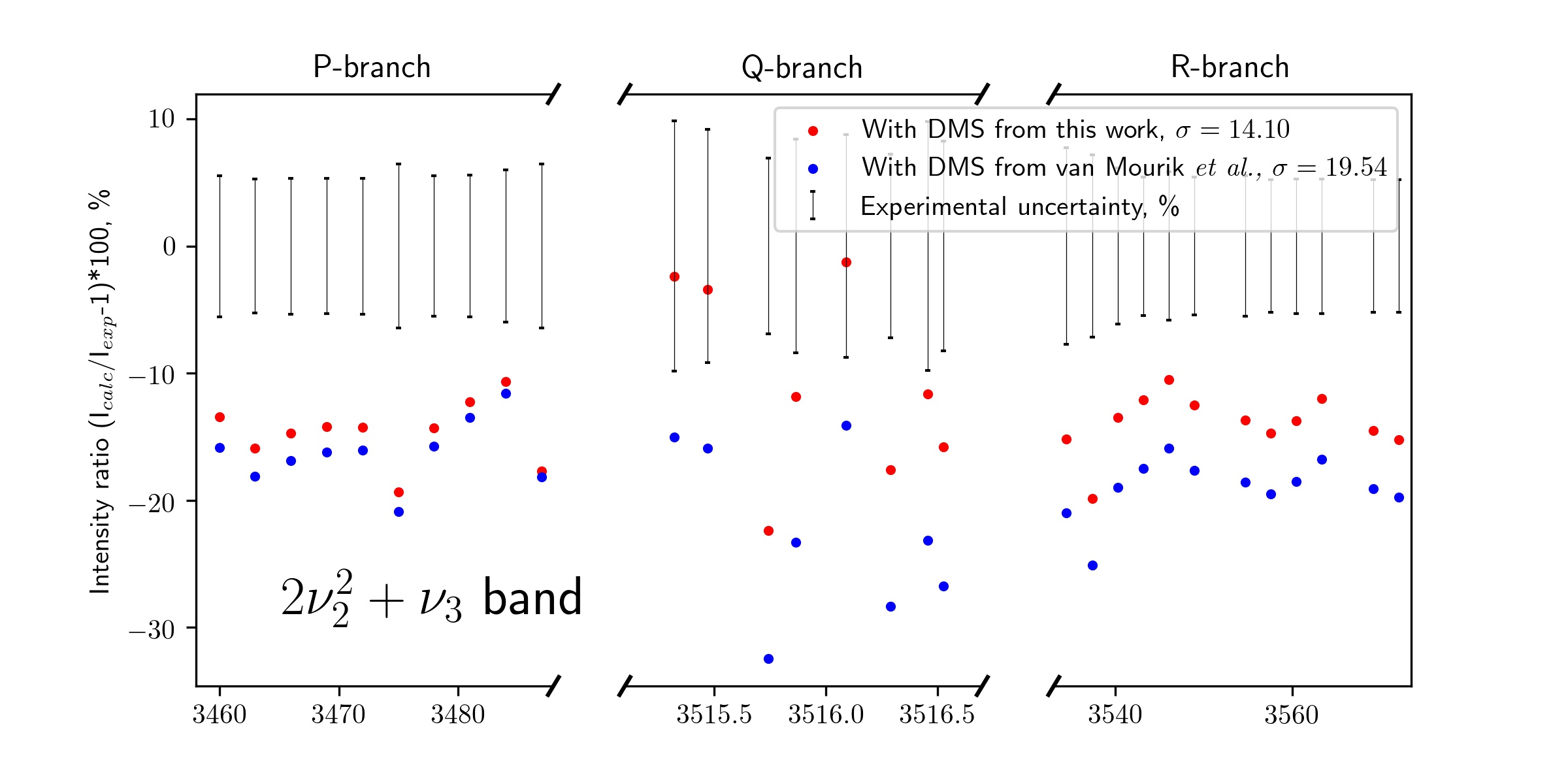}}
%\caption{Comparison of line intensities for the ($0221-0000$)  band.}
%\label{fig:2nu22nu3}
%\end{figure}

\vspace{-1cm}

%\begin{figure}[]
\center{\includegraphics[scale=1.0,trim= 50 0 0 0]{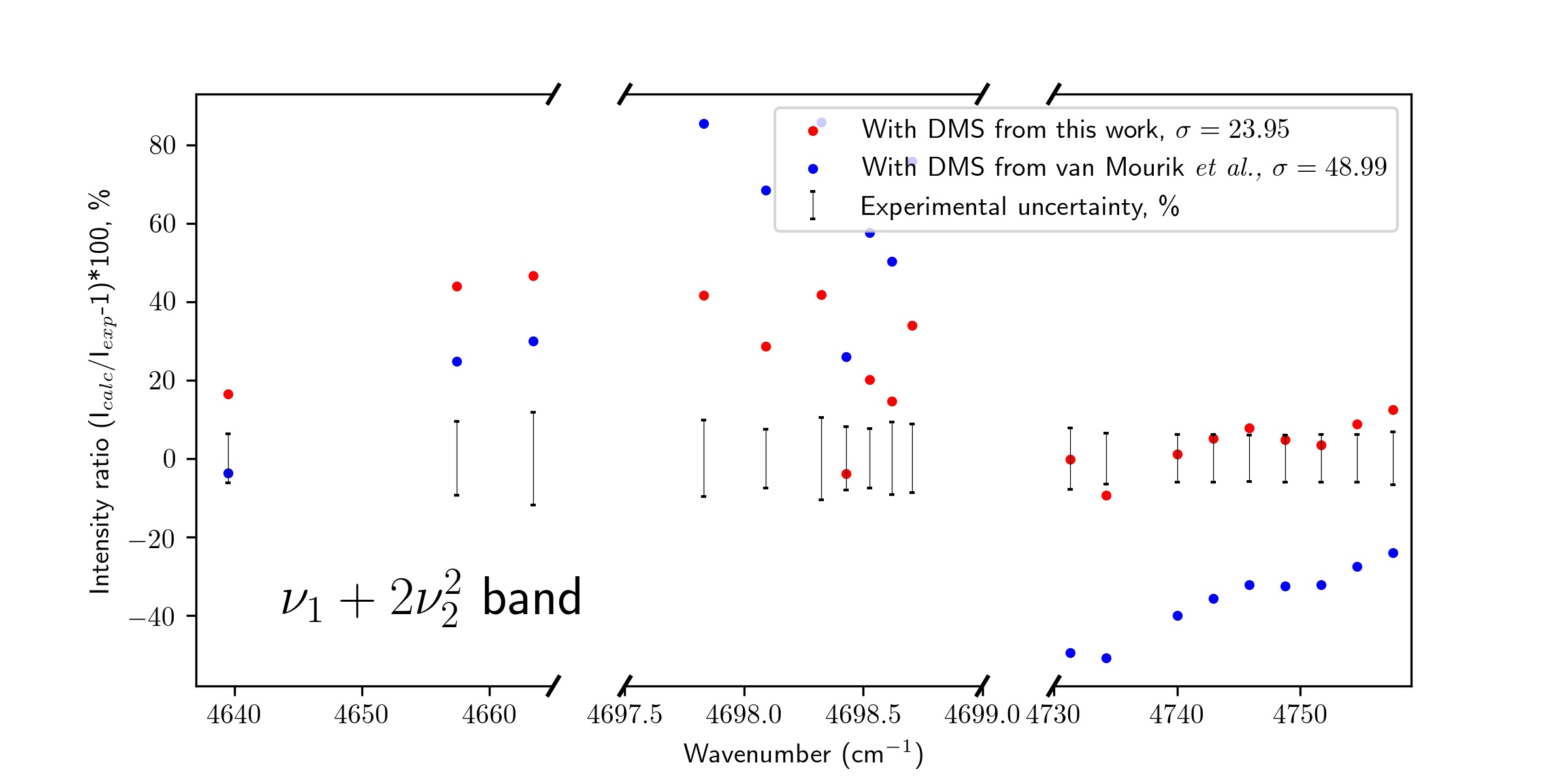}}
\caption{Comparison of line intensities for the ($0221-0000$) (upper) and ($1220-0000$) (lower) bands.}
\label{fig:2nu22nu3}
\end{figure}

While the $2\nu_{2}^{2}$ and $3\nu_{2}^{3}$ pure bending overtone
bands (figure \ref{fig:3nu23}) show no improvement, the
$2\nu_{2}^{2}+\nu_{3}$ combination band is reproduced with an accuracy
similar to the pure CN-stretch lines (figure \ref{fig:2nu22nu3}), and
the corresponding deviation reduces from 20~\% to 14~\%.  The
$\nu_{1}+2\nu_{2}^{2}$ combination band (figure \ref{fig:2nu22nu3})
is predicted  with the ``old global'' and ``new'' DMSs with standard
deviations from experiment of 49 and 24 \%,
respectively.  Such large absolute differences may be caused by the
high sensitivity of these weak lines to the parameters of the
calculation.

\subsubsection{H$^{13}$CN $2\nu_{1}$ (stretching H--C)}
We performed calculations on the H$^{13}$CN isotopologue to compare
with the recent measurements reported by Guay \etal~\cite{nist_h13cn}; 
a comparison is presented in table \ref{tab:nist}.
Our calculations give good agreement with the observed transition intensities
except for the P(20) line. Intensity difference for this line
are around 12~\% while for other lines the corresponding values are below
4~\%.  This unexpected deviation led us to analyze carefully the
$2\nu_1$ wavenumber range.

Hebert \etal~\cite{nist-chip-comb-17} (see figure 7) studied the
P-branch of H$^{13}$CN also studied by Guay \etal\ and showed that the
$2\nu_{1}+\nu_{2}^{1}-\nu_{2}^{1}$ ``hot'' band lies in the same
wavenumber range.  We find that the frequency of the
$2\nu_{1}+\nu_{2}^{1}-\nu_{2}^{1}$ P(9) line differs by less than 0.04
\cm\ from the one of the $2\nu_{1}$ P(20) line. We therefore suggest
that these two lines are blended meaning that the observed absorption
depends on the sum of their intensities.

%\red{Table 8: What does "S" mean? The units of "S"?}
\begin{table}
  \caption{
%\blue{How does $S$ differ from $I$? If they are the same
%      use the same symbols} 
      Comparison of calculated H$^{13}$CN $2\nu_{1}$ line
    intensities ($I$, cm/molecule, powers of 10 in parenthesis) 
    with measurements performed at NIST \cite{nist_h13cn}. Intensity difference presented 
    as $\varepsilon = (I_{\text{calc}} / I_{\text{obs}} - 1) \cdot 100~\%$ is 
    performed with values obtained from ExoMol linelist \cite{jt570}, calculated with 
    the ``old global'' DMS from \cite{jt273} and the fitted PES from this work, and 
    obtained with both new PES and DMS from this work. }
\label{tab:nist}
\begin{tabular}{lcrrr}
\hline\hline
Line	&	$I$	&	$\varepsilon_{\text{ExoMol}}$	&	$\varepsilon_{\text{olddip}}$	&	$\varepsilon_{\text{newdip}}$	\\
\hline
R(8)	&	7.63(-23)	&	4	&	1.5	&	1.8	\\
P(11)	&	5.97(-23)	&	8	&	4.6	&	4.8	\\
P(14)	&	4.58(-23)	&	4	&	0.7	&	0.8	\\
P(16)	&	3.48(-23)	&	1	&	-1.8	&	-1.6	\\
P(17)	&	2.85(-23)	&	3	&	0.5	&	0.7	\\
P(20)	&	1.75(-23)	&	-11$^{a}$	&	-1.7$^{a}$	&	-1.3$^{a}$	\\
P(23)	&	7.2(-24)	&	-1	&	-3.3	&	-3.2	\\
P(24)	&	5.02(-24)	&	6	&	3.6	&	3.7	\\
\hline\hline
\multicolumn{5}{l}{$^{a}$ see comments in text}
\end{tabular}
\end{table}

\section{Conclusion}
\label{sec.diss}

In this work we present a new spectroscopically-determined
potential energy surface based on 
a recent \aipes\ and \ai\ dipole moment surface for HCN, which can be
used for increased accuracy calculations of transition intensities. These 
data may be used in various applications such as the study of astronomical 
objects or intramolecular dynamics.

Spectroscopically-determined PESs are constructed for both isomers of
[H,C,N] system -- HCN and HNC. Experimental energy levels of hydrogen
cyanide are reproduced with a standard deviation $\sigma=$0.0373 \cm,
which is an order of magnitude better than the corresponding value for
the \ai\ surface.  Energy levels for the significantly less harmonic
hydrogen isocyanide molecule are reproduced with $\sigma=0.37$ \cm\
compared to the \ai\ value of $\sigma=4.1$ \cm.
%\blue{How does this number compare with \aipes?}
Differences of a few percent are obtained for many calculated 
transition intensities obtained using the \ai\ and our new, empirical PESs. 
This shows the importance of high quality potentials for accurate transition intensity predictions.

A new \ai\ DMS was created for HCN well to cover all transition
between energy levels in 0-7200 \cm\ range. For all fundamental
transitions we compute intensities with improved agreement with the
experimental data. In particular, the anomalous behaviour of the
$\nu_3$ R-branch transitions observed by Maki
\etal~\cite{95MaQuKl.HCN} is described better than by previous
studies.

We note that Mellau and co-workers provide energy levels which extend
well beyond the 7200 \cm\ cut-off applied here for both HCN 
\cite{08MeWiWi.HCN,11Mexxxb.HCN} and HNC \cite{10Mexxxb.HNC,11Mexxxx.HNC}. 
Attempts by us
to include these levels in our separate-well treatment failed to give
good results.
%\blue{I do not understand the following statement. You do not have
%a global PES so how does this work help a global study?}
It would be desirable to extend the
accuracy of the low-energy HCN and HNC PESs and DMS to a
globally accurate PES and DMS for this system. Such surfaces could
be used in the subsequent
production of a global accurate linelist of this system and to study
behavior in the region of the saddle point between the two minima.  Work
in this direction is currently in progress.

%\begin{acknowledgement}
\section*{Acknowledgement}
This work  was supported by RFBR as part of the research project \# 18-32-00698. 
Authors thank Eamon K. Conway
for help with electronic structure computations. 
MVYu also thanks Dmitry S. Makarov  for helpful discussion during the course of this work.

%\end{acknowledgement}

%\bibliographystyle{model1a-num-names}
\bibliographystyle{elsarticle-num}

%\bibliography{journals_phys,jtj,HCN,CH3CN,hcnfit,PH3,exoplanets,programs,methods,DMS}

\end{document}